\documentstyle[psfig]{l-aa}
\begin{document}
\renewcommand{\topfraction}{1.}
\renewcommand{\bottomfraction}{1.}
\renewcommand{\textfraction}{0.}
\thesaurus{06(08.13.2; 08.16.4; 09.04.1; 11.13.1; 13.09.01; 13.09.6)}
\title{Obscured Asymptotic Giant Branch stars in the Magellanic Clouds
       III. New IRAS counterparts\thanks{based on observations
       collected at the European Southern Observatory, La Silla, Chile
       (proposal ESO 56.E-0681)}}
\author{Jacco Th. van Loon\inst{1,2}, Albert A. Zijlstra\inst{1},
        Patricia A. Whitelock\inst{3}, L.B.F.M. Waters\inst{2}, Cecile
        Loup\inst{4,5}, Norman R. Trams\inst{6}}
\institute{European Southern Observatory, Karl-Schwarzschild
           Stra{\ss}e 2, D-85748 Garching bei M\"{u}nchen, Germany
      \and Astronomical Institute, University of Amsterdam, Kruislaan
           403, NL-1098 SJ Amsterdam, The Netherlands
      \and South African Astronomical Observatory, P.O.Box 9, 7935
           Observatory, South Africa
      \and European Southern Observatory, Casilla 19001, Santiago 19,
           Chile
      \and Institut d'Astrophysique de Paris, 98bis Boulevard Arago,
           F-75014 Paris, France
      \and ISO Science Operations Centre, Astrophysics Division of
           ESA, Villafranca del Castillo, P.O.Box 50727, 28080 Madrid,
           Spain}
\date{Received date; accepted date}
\maketitle
\markboth{Jacco Th.\ van Loon et al.: Obscured AGB stars in the
          Magellanic Clouds III}{Jacco Th.\ van Loon et al.: Obscured
          AGB stars in the Magellanic Clouds III}
\begin{abstract}

We have searched for near-infrared stellar counterparts of IRAS point
sources in the Large Magellanic Cloud (LMC), in J- and K-bands. This
resulted in the detection of 21 counterparts, of which 19 are new
discoveries. Using colour--magnitude and colour--colour diagrams, we
identify 13 Asymptotic Giant Branch (AGB) stars with thick
circumstellar dust envelopes, 7 possible early post-AGB stars or stars
recovering from a thermal pulse, and 1 red supergiant or foreground
star. For 10 of the IRAS targets we do not succeed in detecting and/or
identifying a near-infrared counterpart. We serendipitously detect 14
other red sources, of which 2 are known Long Period Variables, and a
few galaxies. The near-infrared and optical colours of the galaxies
may indicate considerable interstellar extinction through the LMC, as
much as $A_V \sim$ 2--4 mag. The relative number of AGB carbon stars
over oxygen stars is shown to decrease as the luminosity
increases. Yet amongst the faintest mass-losing AGB stars oxygen-rich
stars still exist, which puts constraints on current convection
theories that predict the occurrence of third dredge-up and Hot Bottom
Burning. We investigate the nature of some LMC stars that have
infrared properties very similar to suspected Galactic post-AGB
stars.

\keywords{Stars: mass loss -- Stars: AGB and post-AGB -- dust,
extinction --- Magellanic Clouds -- Infrared: galaxies --- Infrared:
stars}
\end{abstract}

\section{Introduction}

Population studies of Asymptotic Giant Branch (AGB) stars in the
Large Magellanic Cloud (LMC) have led to a number of interesting
results, using the advantage of its known distance. The
period--luminosity relation for Mira variables was deduced from
photometric monitoring of Long Period Variables (LPVs) in the LMC
(Feast et al.\ 1989). Objective prism survey studies of the LMC showed
a clear lack of carbon stars amongst the most luminous AGB stars ($-6
> $M$_{\rm bol} > -7$ mag; see Iben 1981 for a review). These studies
have been largely based on samples of optically visible AGB stars
(e.g.\ Blanco et al.\ 1980; Cohen et al.\ 1981; Westerlund et al.\
1981).

Stars near the tip of the AGB experience mass-loss rates as high as
$10^{-5}$ M$_{\odot}$~yr$^{-1}$, giving rise to circumstellar
envelopes (CSEs) in which dust forms, making them optically thick at
optical wavelengths. Hence these AGB stars may only be observable at
infrared (IR) wavelengths and were missed in earlier studies. Samples
of candidate obscured AGB stars in the LMC have only recently become
available (Reid 1991; Wood et al.\ 1992; Zijlstra et al.\ 1996). The
AGB nature of some of these stars still needs to be confirmed, and
their (circum-)stellar properties are yet to be explored in
detail. The small existing samples should be extended both in number
and in (circum-)stellar parameter space, to allow a population study
of mass-losing AGB stars in the LMC.

In paper~I (Loup et al.\ 1997) we constructed an IRAS-selected sample
of AGB and post-AGB star candidates. In paper~II (Zijlstra et al.\
1996) we confirmed the AGB nature for a subsample thereof, on the
basis of their bolometric luminosities and their near-IR (NIR) and
IRAS colours. We also showed that the chemical type --- oxygen or
carbon star --- may be deduced from the position of a mass-losing AGB
star in the K--[12] versus H--K diagram. Considering the difficulty of
obtaining spectra of these stars, the IR colour diagnostics are a
powerful tool in determining the nature of the IR stars (e.g.\ van der
Veen \& Habing 1988; Guglielmo et al.\ 1993; Le Bertre 1993; Le
Sidaner \& Le Bertre 1994; Loup \& Groenewegen 1994).

Here, in paper~III of our series on the study of obscured AGB stars in
the Magellanic Clouds, we present the results of a NIR search for
counterparts of IRAS point sources in the direction of the LMC, aiming
at enlarging the sample of known obscured AGB stars in the LMC. After
classifying the stars as being mass-losing AGB stars, or not, we
briefly comment on their properties. We discuss not only the AGB
stars, but also the nature of other positive identifications, and
serendipitous detections. Amongst them are stars that may be explained
as being in the post-AGB phase, and we compare their properties with
those of a sample of IRAS-selected post-AGB star candidates in the
Milky Way. We also find a few galaxies, which may be used as probes to
measure the interstellar extinction inside the LMC.

\section{Observations}

\subsection{Sample}

%
%
\begin{figure}[tb]
\centerline{\psfig{figure=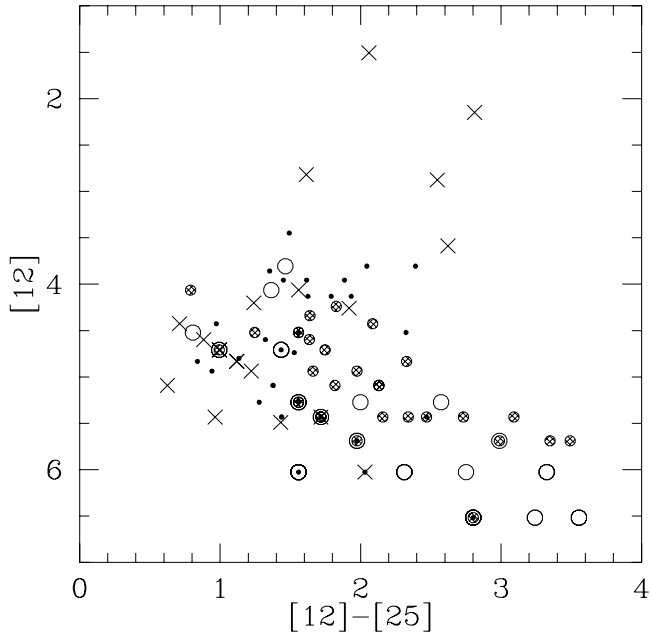,width=90mm}}
\caption[]{IRAS 12~$\mu$m magnitudes versus IRAS [12]--[25] colours
for the sample of IRAS sources that we selected as targets for
observation in the NIR (open circles), the targets that we actually
observed (shaded circles), the AGB star candidates from paper~II
(dots), and the red supergiants from paper~II (crosses)}
\end{figure}

For our search for NIR counterparts, we reduced the list of nearly 200
best-candidate sources from paper~I to a list of 68 sources, excluding
stars that were studied in paper~II (except for one source). Our
selection criteria are either IRAS flux densities $S_{25} > S_{12}$,
or $S_{12} > 0.35$ and $S_{25} > 0.22$ Jy. Limited by the available
observing time, we finally observed 31 sources. They were selected
mainly amongst the sources that were relatively bright at 12~$\mu$m,
had red [12]--[25] colours, and were not affected by cirrus at
60~$\mu$m. We adopt [12]$ = -2.5 \log{(S_{12}/28.3)}$ and [12]--[25]$
= -2.5 \log{(S_{12}/S_{25})\times(6.73/28.3)}$, with $S_{12}$ and
$S_{25}$ the flux density in Jy in the IRAS 12 and 25~$\mu$m bands,
respectively (IRAS Explanatory Supplement 1988). In the IRAS [12]
versus [12]--[25] magnitude--colour diagram shown in Fig.\ 1, these 31
sources are represented by shaded circles, whereas the remaining 37
sources are plotted as open circles. Several sources coincide in this
diagram, because of the discrete values for the IRAS flux densities.
The diagram shows that we mostly selected the brighter and slightly
redder sources. We also plotted the AGB star candidates (dots) and red
supergiants (crosses) from paper~II. The present sample is fainter at
12~$\mu$m and has a redder [12]--[25] colour on average than the AGB
star candidates from paper~II, that are themselves fainter at
12~$\mu$m and have redder [12]--[25] colours than the supergiants.

\subsection{J and K-band imaging photometry}

We observed on the clear nights 1/2 and 2/3 January 1996, using the
NIR camera IRAC2 at the ESO/MPI 2.2m telescope at La Silla,
Chile. Lens C was chosen to cover a field of view of
$133^{\prime\prime} \times 133^{\prime\prime}$, with a pixel scale of
$0.49^{\prime\prime}$. This combines the advantages of a large field
of view and a reasonable sampling of the point spread function for
doing photometry. For each IRAS point source field we did a sequence
of 12 images in the K-band filter, each consisting of ten 3-second
exposures, shifted by $5^{\prime\prime}$ in right ascension with
respect to the previous image. We repeated this procedure backwards,
using the J-band filter. The search is deepest in the $\sim
75^{\prime\prime} \times 130^{\prime\prime}$ centred at the IRAS point
source. We compared the IRAC2 fields with the Digitized Sky Survey
(http://archive.eso.org/dss/dss), to determine the actual field
centres. The absolute pointing of the telescope was found to be
accurate to about $10^{\prime\prime}$, from a comparison of the IRAC2
field centres and the IRAS positions used to point the telescope.

We constructed frames that represent a good approximation of the
background emission, by median-averaging the shifted images within
each sequence, rejecting the brightest pixels to avoid contamination
by stars. The background-subtracted images were flat-field corrected
using flatfields obtained by taking an image of a screen in the dome,
illuminated by a lamp dedicated for this purpose, and subtracting a
similar image with the lamp off. The individual images were shifted to
bring the position of the IRAS point source back in the centre, by
integer number of pixels to ensure flux conservation. Finally these
images were added together. In this way, we could detect stars down to
limiting magnitudes of $J \sim 20$ mag, and $K \sim 18$ mag. The
limiting magnitude varied from field to field by about a magnitude,
mainly due to differences in the background.

The standard stars were observed by taking an image consisting of an
average of thirty 0.6-second exposures, and another image shifted by
$40^{\prime\prime}$. Their difference yields a background-subtracted
frame, which we then flat-field. The standard stars were observed
regularly during the night, covering the same air masses as the LMC.

The J and K-band images were blinked to identify the J--K reddest
sources. On these, circular aperture photometry was done with an
increasing aperture radius, to create a radial magnitude profile of
the star. The same was done for the standard star. From the
differential magnitude profile the magnitude of the program star can
be estimated, as well as its accuracy. The standard stars used are
HD38150 (J$_{\rm SAAO}=8.210$ mag, K$_{\rm SAAO}=7.880$ mag),
SA94--702 ($=$ GSC 00048--00918; J$_{\rm SAAO}=9.246$ mag,
K$_{\rm SAAO}=8.289$ mag), and HD52467 (J$_{\rm SAAO}=8.637$ mag,
K$_{\rm SAAO}=8.699$ mag). These magnitudes are in the SAAO system
(cf.\ Carter 1990), and the IRAC2 magnitudes have been converted to
the SAAO system using relations derived by Lidman (1995) on a single
test night:
\begin{equation}
\left( \begin{array}{c} {\rm J}_{\rm IRAC2} \\ {\rm K}_{\rm IRAC2}
\end{array} \right) = \left( \begin{array}{rr} 0.883 & 0.117 \\ -0.055
& 1.055 \end{array} \right) \times \left( \begin{array}{c}
{\rm J}_{\rm SAAO} \\ {\rm K}_{\rm SAAO} \end{array} \right)
\end{equation}
The 1-$\sigma$ errors are 0.012 and 0.015 on the coefficients for the
J$_{\rm IRAC2}$ and K$_{\rm IRAC2}$ magnitudes, respectively. The
conditions during the observations were photometric. Extinction
corrections were found to be 0.05 mag air-mass$^{-1}$ in the J-band,
and 0.03 mag air-mass$^{-1}$ in the K-band on the first night, and
twice as large on the second night. The relative humidity, dome
temperature, and seeing were about 60---70\%, $13^\circ$C, and
0.7---0.8$^{\prime\prime}$ on the first night, and about 40---60\%,
15---12$^\circ$C, and 0.7---1$^{\prime\prime}$ on the second night.

\subsection{BVRi-band imaging photometry}

We observed on the clear night 24/25, and the partially cloudy night
30/31 December 1996, using the direct imaging camera at the Dutch 0.9m
telescope at La Silla, Chile.

On the first night, we imaged two $3.8^\prime \times 3.8^\prime$
fields centred on the galaxies that we detected in the NIR close to
the IRAS point sources LI--LMC0603 and LI--LMC1818 (see below). The
night was photometric, although the Moon was close to opposition. We
took six images in the Bessel B-band, and three images in the Bessel
R-band, all 200 seconds integration time per frame. Stellar images on
these frames had FWHM of typically
1.1$^{\prime\prime}$--1.4$^{\prime\prime}$. The images were calibrated
by stars in the Landolt (1992) standard star field SA98 that were
observed close in time and airmass to the galaxy fields. Hence the
B and R magnitudes are on the photometric systems of Johnson and
Kron-Cousins, respectively. We obtained integrated magnitudes for the
galaxies and the few redmost stars in the fields, using aperture
photometry and applying aperture corrections derived from the much
less crowded standard star fields.

On the second night, we measured the bright stellar counterpart (see
below) of LI--LMC1821 in the Bessel B, V, and R-bands, and the Gunn
i-band. The night was cloudy, and useful photometry could only be
obtained by rapidly switching back and forth between the star and
$\eta_1$ Dor (= HR2194), an A0 Main Sequence star of V$ = 5.7$ and
B--V$ = -0.03$ mag at 1.5 degree separation, until stable and
well-defined magnitudes could be determined. The B, V, and R magnitudes
that we obtained for LI--LMC1821 are on the Johnson photometric system,
whereas the i magnitude is on the Gunn photometric system. We estimated
the Gunn i-band magnitude for $\eta_1$ Dor to be i$ = 5.74$ mag.
Integration times were 0.3, 0.2, 0.2, and 0.5 seconds for BVRi
respectively. To avoid problems with shutter delay times for these
short integration times, we used the same integration times for both
stars, and they were always positioned at the same place on the CCD.

\section{Results}

\subsection{Identification with IRAS sources}

%
%
\begin{table*}
\caption[]{Names (see text), NIR positions, separations $\Delta$ of
IRAS and NIR source (in arcsec), J- and K-band magnitudes and J--K
colours, IRAS 12- and 25~$\mu$m fluxes (in Jy) and [12]--[25] colours
for the objects in our sample. We adopt [12]--[25]$ = -2.5
\log{(S_{12}/S_{25})\times(6.73/28.3)}$, with $S_{12}$ and $S_{25}$
the flux density in Jy in the IRAS 12 and 25~$\mu$m bands,
respectively (IRAS Explanatory Supplement 1988). 1-$\sigma$ error
estimates are given for the NIR photometry.}
\begin{tabular}{llllllllllllll}
\hline\hline
LI--LMC & IRAS & RA(2000) & Dec(2000) & $\Delta$ & J & $\sigma_{\rm
J}$ & K & $\sigma_{\rm K}$ & J-K & $\sigma_{\rm J-K}$ & S$_{12}$ &
S$_{25}$ & [12]-[25] \\
\hline
\multicolumn{14}{l}{\it positive identifications: AGB stars in the
LMC} \\
0099 & 04518--6852 & 04 51 37.5 & --68 47 32 & 13 & \llap{$>$}19.3 &
0.5 & 16.57 & 0.13 & \llap{$>$}2.7 & 0.6 & 0.37 & 0.22 & 0.99 \\
0109 & & 04 52 19.9 & --67 22 37 & 29 & 11.61 & 0.01 & 9.13 & 0.01 &
2.48 & 0.01 & 0.22 & 0.22 & 1.56 \\
0136 & 04535--6616 & 04 53 45.0 & --66 11 43 &  9 & \llap{$>$}22.6 &
0.4 & 15.55 & 0.07 & \llap{$>$}7.0 & 0.4 & 0.19 & 0.33 & 2.16 \\
0180 & 04552--6536 & 04 55 27.7 & --65 31 06 & 33 & 15.94 & 0.04 &
11.69 & 0.02 & 4.25 & 0.04 & 0.15 & 0.22 & 1.98 \\
0297 & 05003--6712 & 05 00 19.4 & --67 07 53 & 10 & 12.04 & 0.06 &
9.48 & 0.04 & 2.56 & 0.07 & 0.44 & 0.44 & 1.56 \\
0344 & 05026--6809 & 05 02 21.7 & --68 05 22 & 47 & \llap{$>$}20.6 &
0.4 & 12.93 & 0.04 & \llap{$>$}7.7 & 0.4 & 0.33 & 0.67 & 2.33 \\
0603 & 05125--7035 & 05 12 00.4 & --70 32 22 &  4 & \llap{$>$}19.0 &
0.2 & 13.99 & 0.04 & \llap{$>$}5.0 & 0.2 & 0.41 & 0.44 & 1.64 \\
0782 & 05187--7033 & 05 18 09.3 & --70 31 15 & 46 & 15.63 & 0.08 &
11.64 & 0.04 & 3.99 & 0.09 & 0.19 & 0.78 & 3.09 \\
1092 & 05278--6942 & 05 27 23.8 & --69 39 43 &  3 & \llap{$>$}20.2 &
0.4 & 14.26 & 0.04 & \llap{$>$}5.9 & 0.4 & 0.37 & 0.44 & 1.75 \\
1198 & 05306--7032 & 05 30 06.1 & --70 30 40 &  4 & \llap{$>$}19.8 &
1.0 & 15.88 & 0.04 & \llap{$>$}3.9 & 1.0 & 0.30 & 0.33 & 1.66 \\
1813 & 06025--6712 & 06 02 31.3 & --67 12 47 & 10 & \llap{$>$}20.0 &
0.3 & 13.26 & 0.02 & \llap{$>$}6.7 & 0.3 & 0.44 & 0.44 & 1.56 \\
1817 & 06028--6722 & 06 02 44.8 & --67 22 42 & 15 & 18.39 & 0.05 &
12.50 & 0.02 & 5.89 & 0.05 & 0.67 & 0.33 & 0.79 \\
1818 & 06031--7227 & 06 02 07.2 & --72 27 19 & 12 & \llap{$>$}20.5 &
0.3 & 17.30 & 0.17 & \llap{$>$}3.2 & 0.3 & 0.52 & 0.56 & 1.64 \\
\hline
\multicolumn{14}{l}{\it positive identification: LMC red supergiant or
Galactic foreground star} \\
1821 & 06045--6722 & 06 04 25.2 & --67 23 10 &  6 & 6.9 & 0.1 & 6.4 &
0.1 & 0.5 & 0.1 & 0.44 & 0.33 & 1.25 \\
\hline
\multicolumn{14}{l}{\it tentative identifications: post-AGB or thermal
pulse stars in LMC} \\
0326 & 05019--6751 & 05 01 49.0 & --67 47 28 & 20 & 17.19 & 0.07 &
15.74 & 0.06 & 1.45 & 0.09 & 0.26 & 0.44 & 2.13 \\
0530 & & 05 09 40.4 & --69 24 17 &  6 & 13.01 & 0.06 & 11.94 & 0.04 &
1.07 & 0.07 & 0.22 & 0.22 & 1.56 \\
0777 & 05185--6806 & 05 18 28.1 & --68 04 05 & 44 & 17.05 & 0.05 &
14.79 & 0.03 & 2.26 & 0.06 & 0.19 & 0.39 & 2.34 \\
1316 & & 05 33 12.5 & --69 42 33 & 45 & 17.11 & 0.07 & 14.89 & 0.03 &
2.22 & 0.08 & 0.19 & 0.22 & 1.72 \\
1624 & 05439--6555 & 05 44 07.0 & --65 53 47 & 27 & 19.13 & 0.14 &
16.63 & 0.07 & 2.50 & 0.16 & 0.19 & 0.22 & 1.72 \\
1721 & 05478--7045 & 05 47 12.5 & --70 44 16 & 10 & 17.80 & 0.05 &
15.09 & 0.04 & 2.71 & 0.06 & 0.26 & 0.44 & 2.13 \\
1803 & 05588--6944 & 05 58 26.4 & --69 43 35 & 51 & 18.92 & 0.06 &
16.57 & 0.04 & 2.35 & 0.07 & 0.19 & 0.56 & 2.73 \\
\hline
\multicolumn{14}{l}{\it NIR non-detections} \\
0374 & 05039--7002 & & & & & & & & & & 0.15 & 0.56 & 2.99 \\
0671 & 05150--6942 & & & & & & & & & & 0.30 & 0.44 & 1.98 \\
0770 &             & & & & & & & & & & 0.26 & 0.33 & 1.82 \\
0918 & 05232--7111 & & & & & & & & & &      & 0.22 &      \\
0937 & 05237--7000 & & & & & & & & & & 0.19 & 0.44 & 2.47 \\
1200 &             & & & & & & & & & & 0.57 & 0.73 & 1.83 \\
1232 & 05315--7145 & & & & & & & & & & 0.15 & 0.78 & 3.35 \\
1745 & 05495--7034 & & & & & & & & & & 0.15 & 0.89 & 3.49 \\
1759 & 05509--6956 & & & & & & & & & & 0.48 & 0.78 & 2.09 \\
1768 & 05522--7120 & & & & & & & & & & 0.19 & 0.22 & 1.72 \\
\hline
\multicolumn{14}{l}{\it field stars} \\
0297b & & 05 00 20.1 & --67 08 47 & 45 & 16.52 & 0.07 & 14.98 & 0.06 &
1.54 & 0.09 & & & \\
0603b & & 05 12 04.5 & --70 33 14 & 53 & 16.43 & 0.07 & 14.80 & 0.06 &
1.63 & 0.09 & & & \\
0937b & & 05 23 26.3 & --69 58 49 & 71 & 11.67 & 0.04 & 10.38 & 0.02 &
1.29 & 0.04 & & & \\
1198b & & 05 30 08.0 & --70 30 47 &  8 & 18.45 & 0.27 & 16.81 & 0.18 &
1.6  & 0.3  & & & \\
1624b & & 05 44 01.8 & --65 54 42 & 42 & 18.09 & 0.06 & 16.24 & 0.05 &
1.85 & 0.08 & & & \\
1721b & & 05 47 11.5 & --70 43 50 & 33 & 12.65 & 0.04 & 12.55 & 0.02 &
0.10 & 0.04 & & & \\
1721c & & 05 47 12.4 & --70 44 02 & 20 & 13.98 & 0.04 & 12.82 & 0.02 &
1.16 & 0.04 & & & \\
1721d & & 05 47 11.4 & --70 43 59 & 26 & 15.78 & 0.04 & 15.26 & 0.04 &
0.52 & 0.06 & & & \\
1818b & & 06 02 11.8 & --72 26 41 & 42 & 17.73 & 0.05 & 16.02 & 0.04 &
1.71 & 0.06 & & & \\
1818c & & 06 02 14.5 & --72 27 43 & 30 & 17.77 & 0.04 & 15.83 & 0.06 &
1.94 & 0.07 & & & \\
\hline
\multicolumn{14}{l}{\it galaxies} \\
0603c & & 05 12 05.4 & --70 32 04 & 31 & 15.44 & 0.06 & 13.90 & 0.04 &
1.54 & 0.07 & & & \\
1759b & & 05 50 38.5 & --69 55 41 & 59 & 15.40 & 0.05 & 14.44 & 0.04 &
0.96 & 0.06 & & & \\
1803b & & 05 58 17.3 & --69 44 11 & 41 & 17.56 & 0.04 & 15.61 & 0.03 &
1.95 & 0.05 & & & \\
1818d & & 06 02 12.8 & --72 27 36 & 19 & 15.67 & 0.05 & 13.99 & 0.02 &
1.68 & 0.05 & & & \\
\hline
\end{tabular}
\end{table*}

The positive and tentative identifications, non-detections,
serendipitous detections, and galaxies are listed in Table 1, together
with the IRAS flux densities and NIR magnitudes, where applicable. The
numbers of the stars refer to the LI--LMC numbers in the Schwering \&
Israel (1989, 1990) catalogue of IRAS point sources in the direction
of the LMC. A suffix b, c, or d was added in the case of field stars
or galaxies, neither of which were considered to be the counterpart of
the quoted LI--LMC source. Positive and tentative identifications were
found by first selecting the object in the field of view with the
reddest J--K colour. Stars with J--K$>2$ mag were considered to be the
counterpart of the IRAS source, as no two such red objects ever
appeared in the same field of view. In the absence of objects with
J--K$>2$ mag, objects were considered to be the counterpart of the
IRAS source, if they either had J--K$>1$ mag (i.e.\ cool and/or
reddened stars, see e.g.\ Bell 1992) or were very bright (e.g.\
LI--LMC1821), and not too distant from the position of the IRAS source
(i.e.\ within $\sim 20^{\prime\prime}$). The counterparts with IR
colours similar to those of mass-losing AGB stars have been enumerated
under positive identifications, as this can be considered a
confirmation of their identification with the IRAS source. The other
counterparts have been enumerated under tentative identifications, and
probably are stars with detached CSEs.

LI--LMC0109 was first detected in November 1995, using the NIR
photometer at SAAO. IRAS05003--6712 was already known to be a
mass-losing AGB star in the LMC (paper~II). LI--LMC0530 is identified
with SHV0510004--692755, an LPV with a period of 169 days (Hughes \&
Wood 1990, who measured J$=12.63$ mag and K$=11.51$ mag). We identify
LI--LMC1721b with SHV0547489--704450, an LPV with a period of 264 days
(Hughes \& Wood 1990, who measured J$=12.86$ mag and K$=12.69$ mag):
LI--LMC1721 is not associated with this LPV, because the NIR
counterpart that we detect for this IRAS source is much redder than
the LPV. The same is true for LI--LMC1721c, whereas LI--LMC1721d is
too faint to be the LPV. We identify LI--LMC0937b with
SHV0523536--700128, an LPV with a period of 381 days (Hughes \& Wood
1990, who measured J$=10.84$ mag and K$=9.48$ mag). The IRAS source
LI--LMC1759 lies at the edge of a small open cluster of perhaps a
dozen stars within $0.5^{\prime}$ diameter, and may therefore not be
related to any individual star. The galaxies are discussed below.

\subsection{Serendipitous detections}

%
%
\begin{figure}[tb]
\centerline{\psfig{figure=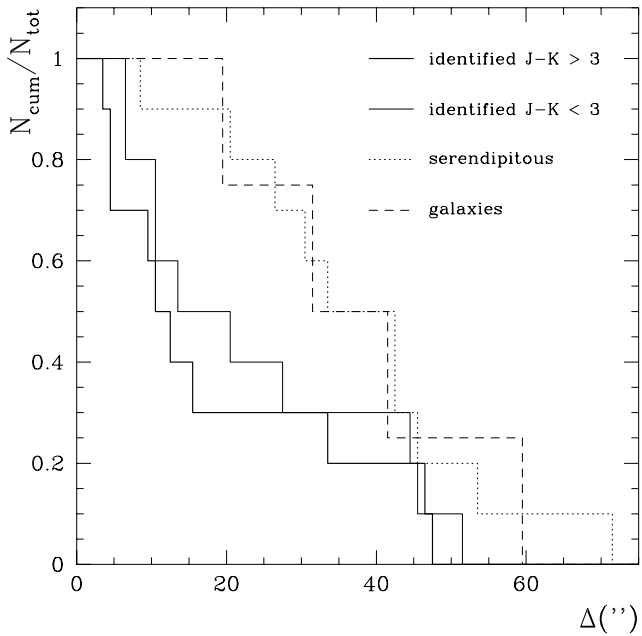,width=90mm}}
\caption[]{Cumulative distribution over separation of the IRAC2 and
IRAS positions, of positive and tentative IRAC2 identifications with
H--K$>3$ (bold solid) and H--K$<3$ (thin solid), serendipitously
detected NIR point sources (dotted) and galaxies (dashed)}
\end{figure}

We have calculated the separations of the IRAS point sources and
the NIR sources detected with IRAC2 (Fig.\ 2: accumulated starting at
infinity, and normalised to the total number of sources). The IRAS
position is generally accurate to 5--15 arcsec, depending mainly on
the 12~$\mu$m flux level, but can be off by more than half an
arc-minute (paper~II).

The reddest sources with J--K$>3$ (bold line) are undoubtedly the
IRAS counterparts. The combined area of our 31 fields covers $\sim
10^{-3}$ square degrees. The area of the LMC covered by the selection
in paper~I measures $\sim 10^2$ square degrees. Only if there were of
the order of $10^5$ such NIR-red sources would one expect to have
detected one serendipitously. There are of the order of $10^2$ IRAS
detected (post-)AGB star candidates (paper~I). Hence the population of
stars with J--K$>3$ that are not detected by IRAS would have to be at
least $10^3$ as many as the IRAS-detected stars in the LMC. The number
of planetary nebulae (PNe) in the LMC is estimated at 1100 (Pottasch
1984). The PN lifetimes of $\sim 10^4$ years are very similar to the
lifetimes of the mass-losing AGB phase (less than $10^5$
years). Assuming that all mass-losing AGB stars will eventually form
PNe, we estimate the number of mass-losing AGB stars in the LMC to be
$10^4$ at most, arguing against the possibility of a population of
$10^5$ LMC stars with J--K$>3$.

The bluer sources (J--K$<3$) that we consider as positive or tentative
identifications (thin line in Fig.\ 2) could be contaminated by NIR
stars that are not associated with the IRAS source. But these cannot
be many, as seen from the similarity of the separation distribution to
that of the reddest sources. The serendipitously detected NIR sources
(field stars) and the galaxies have very similar separation
distributions, much broader than the separation distributions of the
positive identifications. This is indicative of their not being the
IRAS counterpart.

A field of view of $130^{\prime\prime} \times 75^{\prime\prime}$
(the deep area only) implies an expected mean separation of
$40.6^{\prime\prime}$ for a serendipitous detection. The mean
separation for our serendipitous detections is $37^{\prime\prime}$,
and $38^{\prime\prime}$ for the galaxies. We suspect that the
detection probability near the edges is somewhat lower. Essentially
all our detections are situated in the deep area. The mean separation
is $18^{\prime\prime}$ for the reddest, and $23^{\prime\prime}$ for
the bluer positive and tentative identifications. Replacing 2 or 3 of
the reddest stars with separations of $18^{\prime\prime}$ by the same
number of stars but with separations of $37.5^{\prime\prime}$, we
generate a separation distribution with a mean of
$\sim23^{\prime\prime}$. Hence we expect there may be 2 or 3
serendipitous detections in the group of positive and, more likely,
tentative identifications with J--K$<3$. This is too small to affect
the conclusions that we reach in the present study.

\subsection{Galaxies as a probe of the interstellar extinction inside
the LMC}

%
%
\begin{figure*}[tb]
\centerline{\hbox{
\psfig{figure=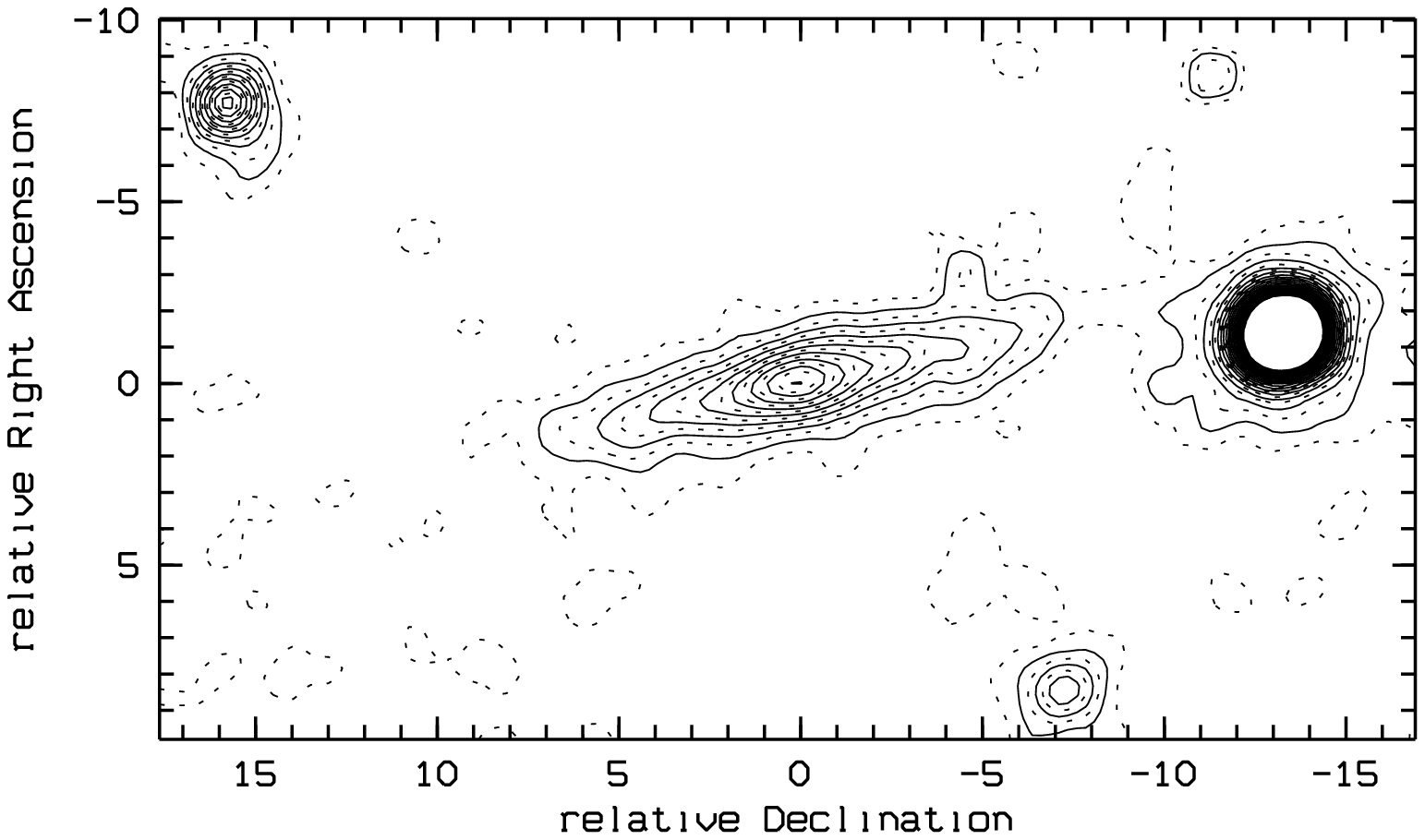,angle=270,width=90mm}
\psfig{figure=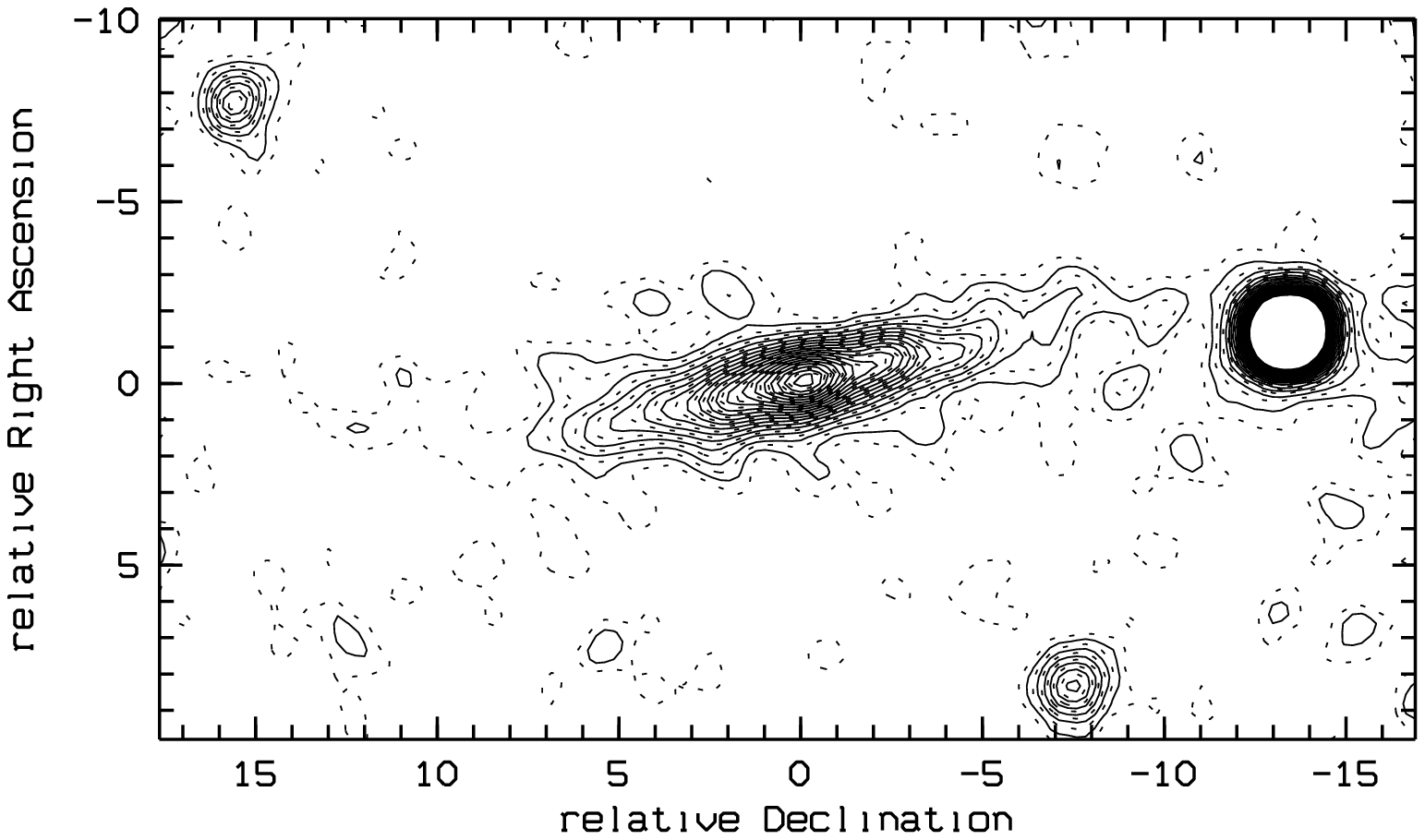,angle=270,width=90mm}
}}
\caption[]{J-band (left) and K-band (right) images of the edge-on
spiral galaxy LI--LMC1818d. Coordinates are in arc-seconds, relative to
the nucleus of the galaxy. Contour levels are between $2.2 \times
10^{-28}$---$1.1 \times 10^{-26}$ erg~cm$^{-2}$~s$^{-1}$~Hz$^{-1}$ per
square arc-second, with intervals of $2.2 \times 10^{-28}$
erg~cm$^{-2}$~s$^{-1}$~Hz$^{-1}$ per square arc-second, the same for
both images. Note that East is down and North is to the left}
\end{figure*}

%
%
\begin{figure*}[tb]
\centerline{\hbox{
\psfig{figure=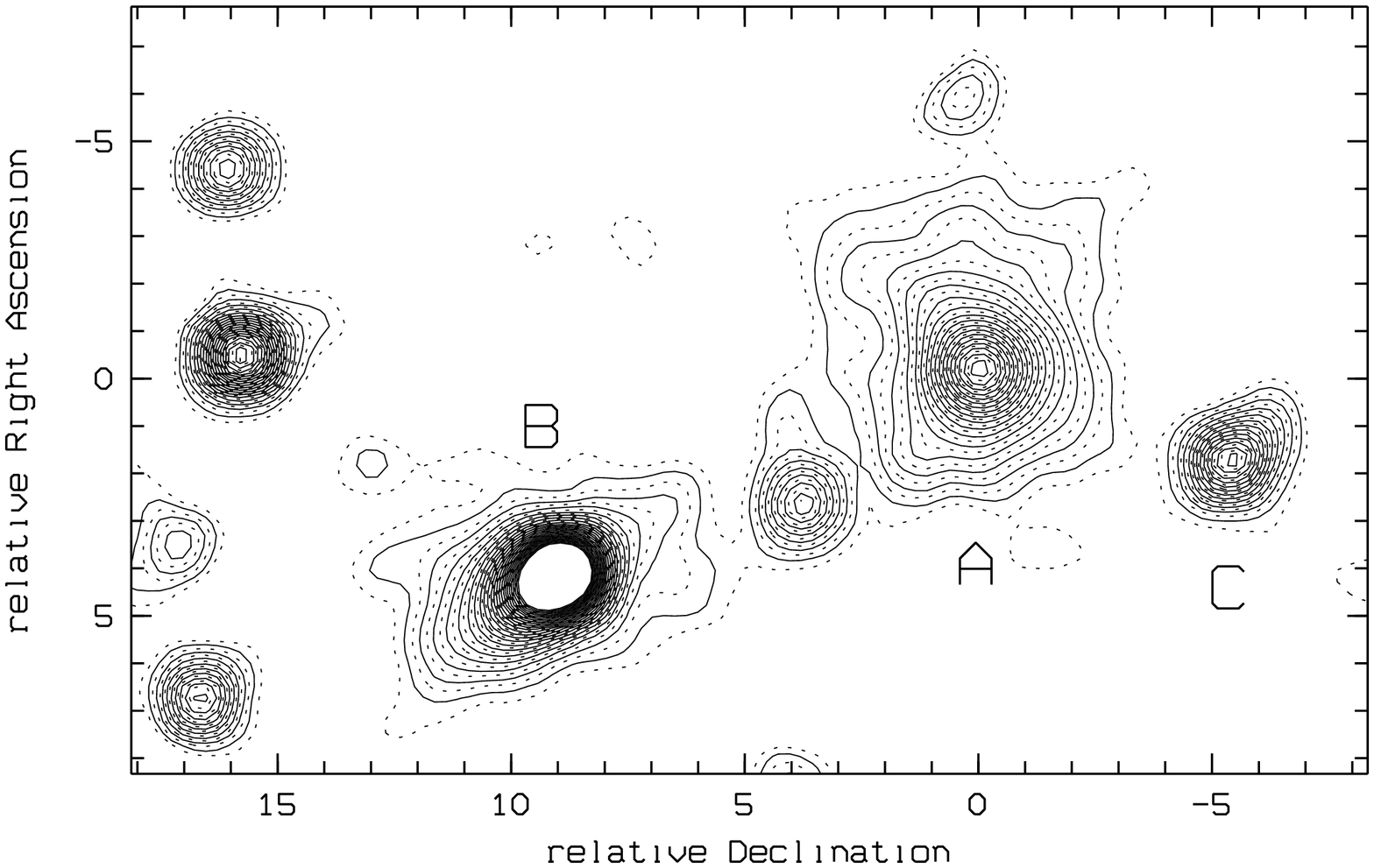,angle=270,width=90mm}
\psfig{figure=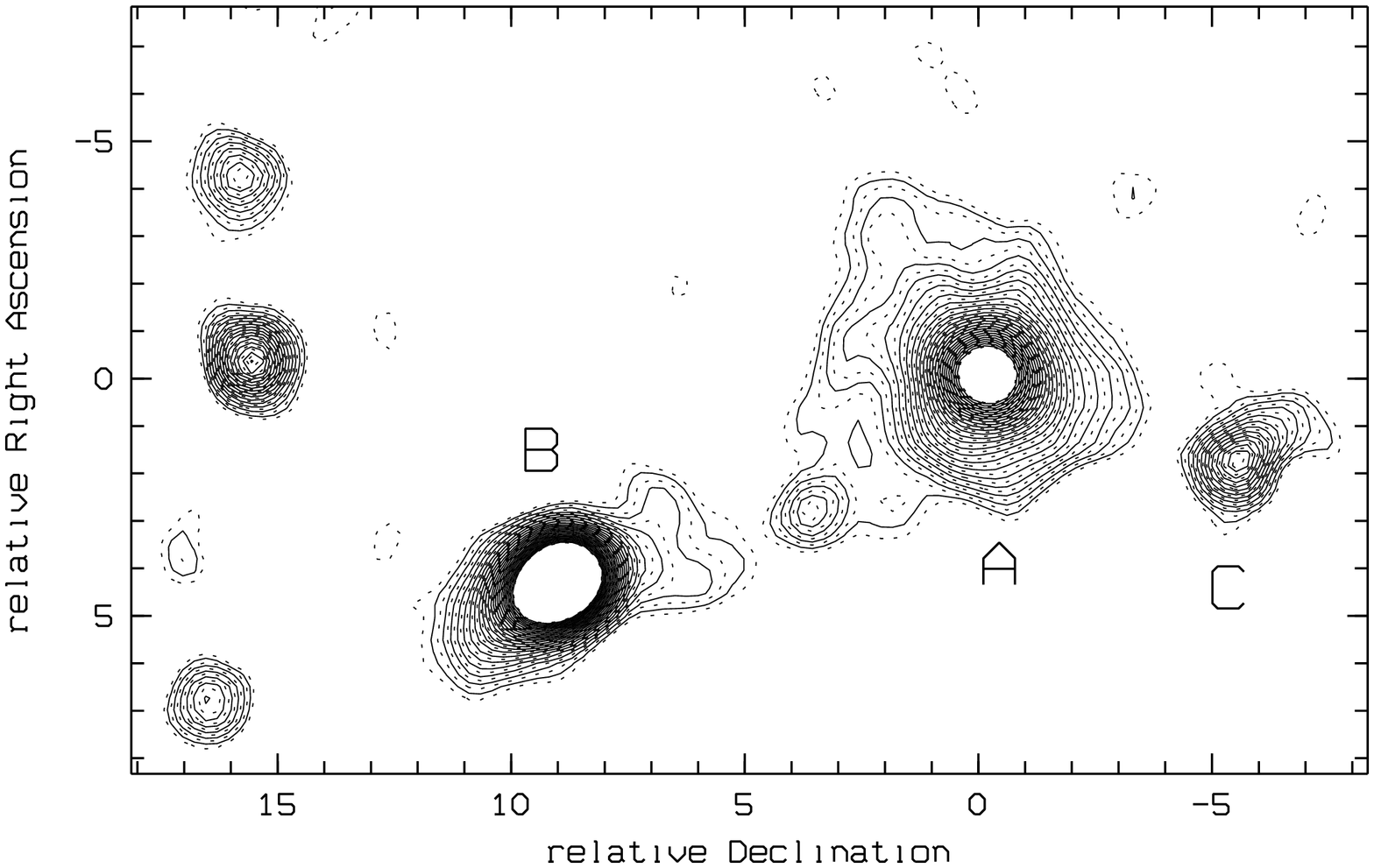,angle=270,width=90mm}
}}
\caption[]{J-band (left) and K-band (right) images of the face-on
spiral galaxy LI--LMC0603c (A), interacting with a second (B), and
possibly a third galaxy (C). Coordinates are in arc-seconds, relative
to the nucleus of galaxy A. Contour levels are between $2.6 \times
10^{-28}$---$6.4 \times 10^{-27}$ erg~cm$^{-2}$~s$^{-1}$~Hz$^{-1}$ per
square arc-second, with intervals of $1.3 \times 10^{-28}$
erg~cm$^{-2}$~s$^{-1}$~Hz$^{-1}$ per square arc-second, the same for
both images. Note that East is down and North is to the left}
\end{figure*}

We discovered a few galaxies, which are probably not related to the
IRAS point sources. Since the galaxies are located behind the LMC,
they could, in principle, be used as probes of the interstellar
reddening through the LMC. If the J--K colour excess of $\sim 0.7$ mag
is due to extinction by dust inside of the LMC, a visual extinction of
A$_{\rm V} \sim 4$ mag is indicated; the extinction through the entire
LMC would vary between A$_{\rm V} \sim 0$ and A$_{\rm V} > 4$ mag. This
could have severe consequences for the observation of stars inside of
the LMC. Although Oestreicher et al.\ (1995) showed the foreground
reddening towards the LMC to be only A$_{\rm V} \sim 0.18$ on average,
stars inside the LMC could suffer from visual extinction of a few
magnitudes.

Are the measured colours of the galaxies intrinsic to them, or have
they been severely affected by interstellar reddening through the LMC?
The intrinsic colours of a galaxy depend on its kind. The magnitudes
for the galaxies approximately represent the total integrated
light. There is no sign of steep colour gradients, although the bulges
or nuclei appear somewhat redder. LI--LMC1818d (Fig.\ 3) is a bright
edge-on spiral galaxy. LI--LMC0603c (Fig.\ 4) is a face-on spiral
galaxy (A) interacting with another galaxy (B), and possibly with a
third galaxy (C). All other objects in the images of LI--LMC1818d and
LI-LMC0603c are unresolved. LI--LMC1759d is probably a spiral galaxy
seen under a small inclination angle, but its position close to the
edge of the field severely degraded the quality of its
image. LI--LMC1803b is small and barely resolved.

Normal galaxies are confined to a colour J--K $\sim$ 0.8---0.9 mag
(e.g.\ Glass 1984; Silva 1996). Only LI--LMC1759d has a colour
consistent with a normal galaxy. LI--LMC1818d may suffer from
extinction by dust in its disk because of its edge-on orientation, but
the red colour of the near-face-on LI--LMC0603c cannot be explained
easily by extinction inside of a normal galaxy.

Galaxies with anomalous IR colours include emission line
galaxies. Whitelock (1985b) presented NIR data on a sample of thirteen
IRAS galaxies, probably H~{\sc ii} galaxies. Their mean J--K colour
was 1.25 mag, with a standard deviation of only 0.13 mag. But the J--K
colours of our reddened galaxies are still consistent with the reddest
H~{\sc ii} galaxies, while both Seyfert 1 and 2 galaxies can reach
J--K colours in excess of 2 mag (Glass \& Moorwood 1985; Almudena et
al.\ 1996; Kotilainen \& Ward 1994). Active galaxies usually have
strong colour differences between the nucleus and the rest of the
galaxy, something which we do not see in our galaxies.

There is less doubt about the question whether our galaxies could be
the counterparts of the IRAS sources in the field. The large distances
between the galaxies and the IRAS positions and/or the presence of a
NIR-redder point source in the field already suggest that the
detection of the galaxies was a mere coincidence. Also, the mid-IR
colours of galaxies detected by IRAS are different from those of our
sources. The IRAS galaxies discussed by Whitelock (1985b) have IRAS
colours typical of cold dust, with S$_{60}$/S$_{25}$ ratios $\sim 8
\pm 1$. This would have lead to 60~$\mu$m flux densities for our IRAS
sources of $\sim 3$---6 Jy, whereas the measurements indicate they
can only be $\sim 1$ Jy at most. Whitelock's galaxies have K
magnitudes of $\sim 11$---12 mag, which is considerably brighter than
our galaxies, whereas their IRAS 25~$\mu$m flux densities are
comparable. Hence we cannot exclude the possibility that our reddened
galaxies have mid-IR excess emission below the sensitivity of IRAS,
but they are not the counterparts of the IRAS sources.

%
%
\begin{table*}
\caption[]{Positions, R-band magnitudes, and B--R colours for the red
sources in the Dutch telescope fields of LI--LMC0603 and LI--LMC1818.}
\begin{tabular}{llllll}
\hline\hline
LI--LMC & RA(2000) & Dec(2000) & R & B--R & remarks \\
\hline
0603b & 05 12 04.5 & -70 33 14 & 18.86 $\pm$ 0.04 & 3.4 $\pm$ 0.9 &
NIR-red star \\
0603c & 05 12 05.4 & -70 32 04 & 17.50 $\pm$ 0.02 & 2.30 $\pm$ 0.12 &
face-on NIR galaxy \\
0603d & 05 11 51.5 & -70 33 32 & 15.45 $\pm$ 0.02 & 5.06 $\pm$ 0.17 &
red star outside IRAC2 field \\
0603e & 05 12 23.4 & -70 33 05 & 15.26 $\pm$ 0.02 & 5.01 $\pm$ 0.11 &
red star outside IRAC2 field \\
\hline
1818b & 06 02 11.8 & -72 26 41 & 19.49 $\pm$ 0.06 & 3.7 $\pm$ 1.5 &
NIR-red star \\
1818c & 06 02 14.5 & -72 27 43 & 19.62 $\pm$ 0.08 & 2.7 $\pm$ 0.4 &
NIR-red star \\
1818d & 06 02 12.8 & -72 27 36 & 17.77 $\pm$ 0.03 & 2.44 $\pm$ 0.24 &
edge-on NIR galaxy \\
1818e & 06 01 54.0 & -72 27 04 & 16.78 $\pm$ 0.02 & 2.47 $\pm$ 0.10 &
galaxy outside IRAC2 field \\
\hline
\end{tabular}
\end{table*}

The B and R-band photometry for the red sources in the fields of
LI--LMC0603 and LI--LMC1818 is presented in Table 2. The stars that we
identified as the NIR counterparts of the IRAS point sources
LI--LMC0603 and LI--LMC1818 were not detected down to $\sim 22^{\rm
nd}$ mag in the R-band. The other NIR-red objects do indeed have
optically red counterparts, both the stars and the galaxies. Moreover,
we discovered two very red stars in the Dutch telescope field of
LI--LMC0603c, and a red galaxy in the Dutch telescope field of
LI--LMC1818d, all three which are outside the corresponding IRAC2
fields. The B--R colour of $\sim 5$ mag for the two red stars
LI--LMC0603d and LI--LMC0603e implies severe inter- or circumstellar
extinction (Whitelock et al.\ 1996). The red B--R colour of galaxy
LI--LMC1818e is remarkably similar to that of the galaxy LI--LMC1818d.
Active galaxies are often optically blue (V\'{e}ron-Cetty \& V\'{e}ron
1996 adopt B--R = 0.57). They can be optically red, though, as is the
case for IRAS galaxies (P.\ V\'{e}ron, private communication; see also
Duc et al.\ 1997).

We have taken the Lauberts-Valentijn ESO catalogue of galaxies
(Lauberts \& Valentijn 1989) to investigate the variation in galaxy
B--R colours across the sky in the vicinity of the LMC. We had to
sample in four square degree bins in order to obtain useful
statistics. We identified a region of 64 square degrees centred at
$4^h$ Right Ascension and $-62^\circ$ Declination, that was relatively
well populated (148 galaxies), and that appeared to be representative
for the colours of galaxies unaffected by extinction through the
Magellanic Clouds system. This region yielded a mean B--R$ = 1.09 \pm
0.17$ mag, with minimum and maximum B--R of 0.70 and 1.38 mag
respectively. We can compare this to the bins containing our galaxies
LI--LMC0603c and LI--LMC1818d. Each of these bins contained one galaxy
from the above mentioned catalogue. Although this is statistically
very poor, it is interesting to note that the galaxy near LI--LMC0603c
has B--R = 1.74 mag, and the galaxy near LI--LMC1818d has B--R = 1.44
mag, both considerably redder than the estimated typical
colour. Comparing the B--R colours of the galaxies LI--LMC0603c and
LI--LMC1818d with the canonical value of B--R$\sim 1.09$, we arrive at
extinctions of $A_V \sim 2.2$ mag.

If we compare the positions of our galaxies with the dust column
density maps by Schwering (1989), we notice that the dust distribution
in the LMC is rather patchy as compared to the SMC. Two of our
reddened galaxies are situated at the East side. Although they lie
outside of the area covered by the dust map, it is at the East side
that there appears to be a massive dust complex. Some of the
serendipitously discovered red stars may thus be normal stars,
situated behind a local, but large amount of dust in the
LMC. Alternatively, they may be the nuclei of normal galaxies as seen
through the LMC. In that case, the dust causing the reddening of the
galaxy colours may be situated behind the main body of the LMC,
leaving the colours of the stars in the LMC unaffected.

The colours and morphologies of the galaxies, the fact that we detect
only a few galaxies of which most are red, and the presence of other
red objects in their projected vicinities all provide circumstantial
evidence for severe interstellar extinction inside or behind the
LMC. However, it is based on very little data, and should be
confirmed by the systematic study of the colours of galaxies seen
through the LMC.

\section{The nature of the IRAS counterparts}

%
%
\begin{figure*}[tb]
\centerline{\hbox{
\psfig{figure=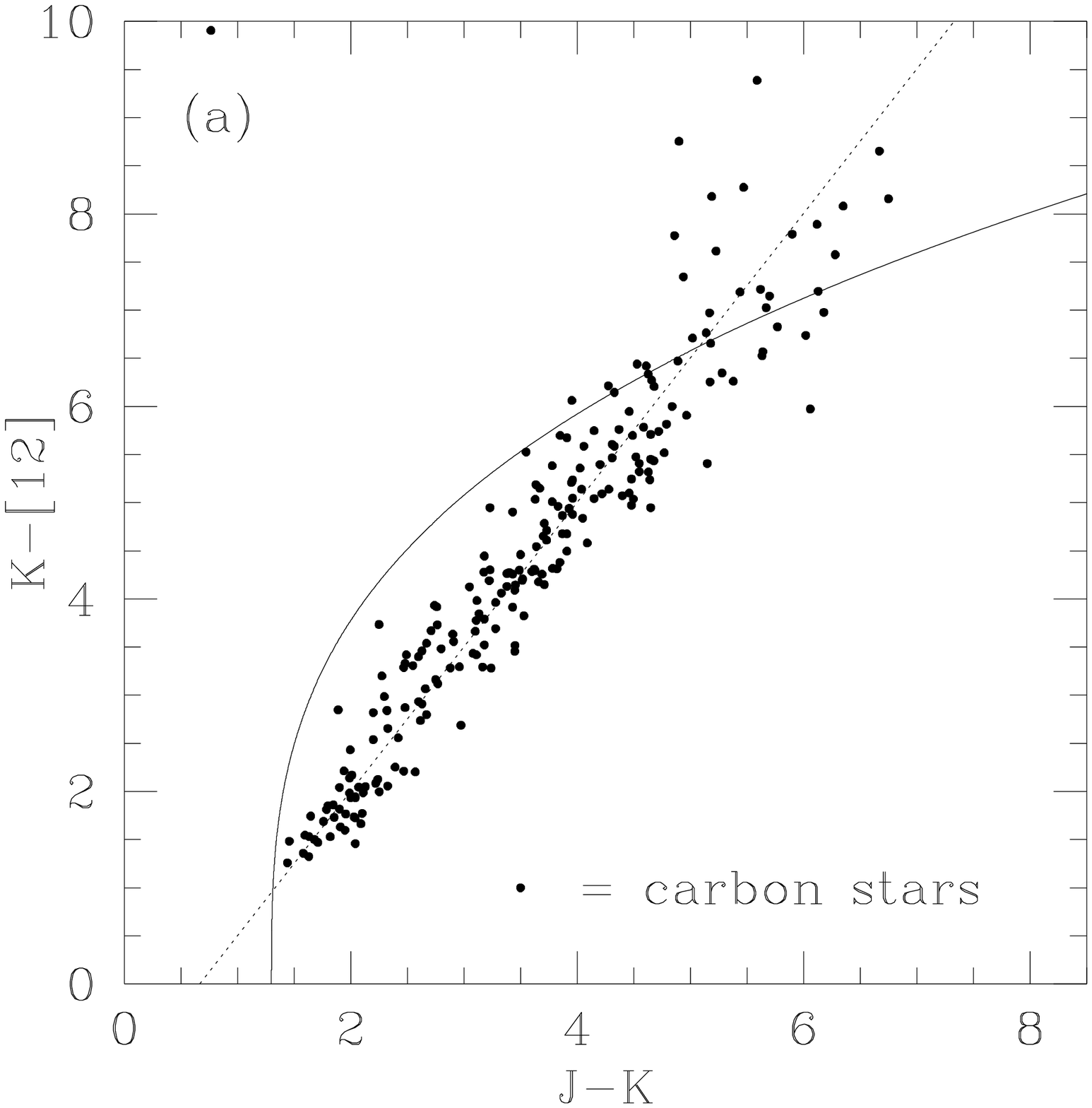,width=90mm}
\psfig{figure=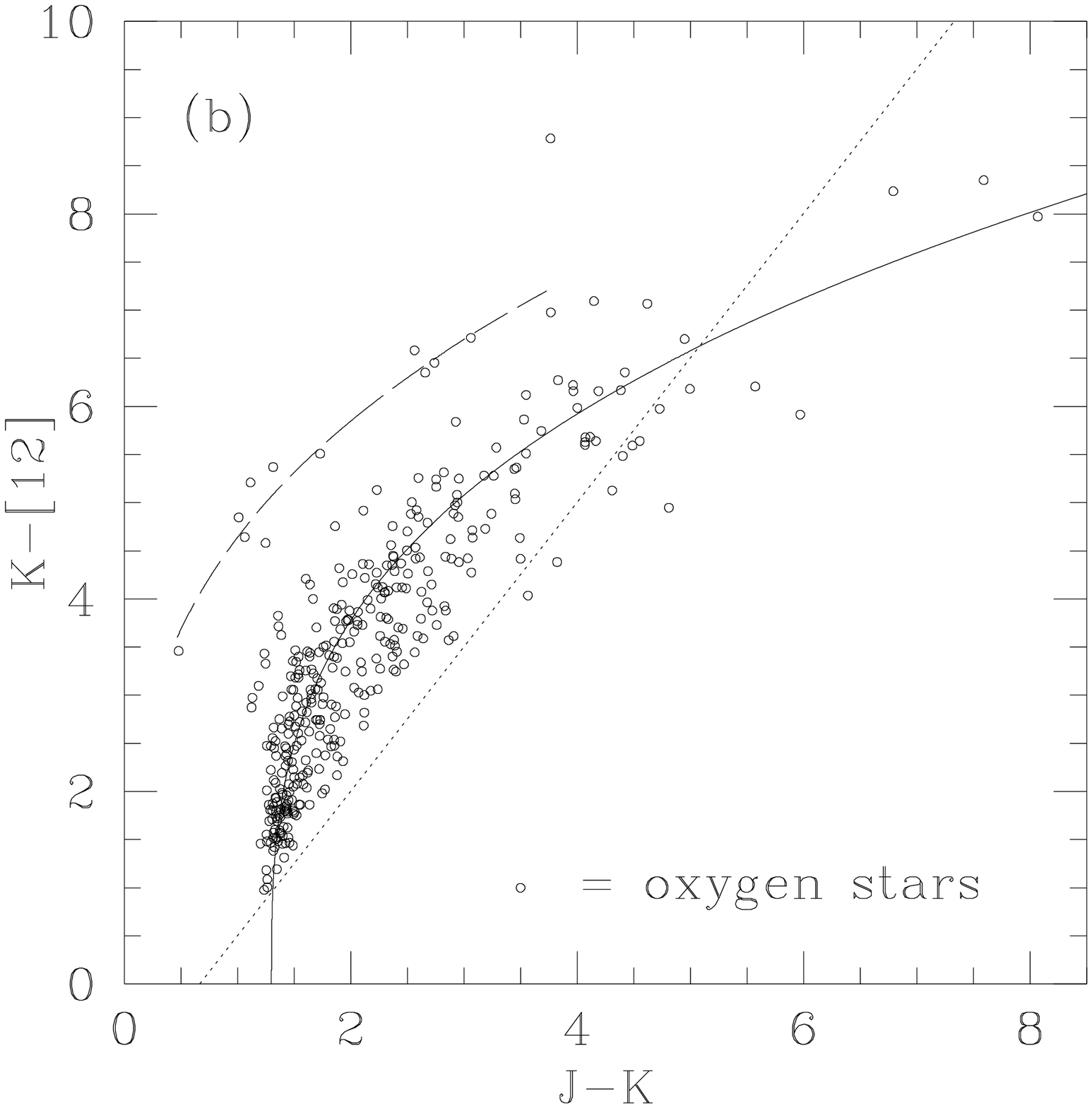,width=90mm}
}}
\caption[]{K--[12] versus J--K diagram for the carbon stars (a) and
oxygen stars (b) of the Galactic sample of Guglielmo et al.\
(1993). We adopt [12]$ = -2.5 \log{(S_{12}/28.3)}$, with $S_{12}$ the
flux density in Jy in the IRAS 12~$\mu$m band (IRAS Explanatory
Supplement 1988). From these data, we derived the carbon star (dotted)
and oxygen star (solid) sequences. The dashed line indicates the
possible existence of a secondary sequence for the oxygen stars}
\end{figure*}

In this section, we classify the IRAS counterparts as mass-losing AGB
stars that are oxygen rich (O), carbon rich (C), or that are not
distinguishable between oxygen stars and carbon stars (OC). We argue
that some of the stars may be post-AGB or thermal pulse stars (post/TP).
The arguments used in classifying the stars are explained below, and
the results are tabulated in Table 4.

\subsection{Chemical classification from colours and magnitudes}

\subsubsection{Galactic comparison sample}

In paper~II we show how in a K--[12] versus H--K diagram carbon
stars are distinguished from oxygen stars. A similar diagram can be
made using J--K. In Fig.\ 5 we present the K--[12] versus J--K
diagnostic diagram, for the carbon stars (Fig.\ 5a) and the oxygen
stars (Fig.\ 5b) from the Galactic sample described in paper~II. It is
clear that in the Milky Way the K--[12] versus J--K diagram can also
be used to separate carbon stars from oxygen stars. The Galactic
carbon star sequence (dotted straight line) is described by the
empirical relation
\begin{equation}
{\rm J-K} = \frac{2}{3} \left( 1 + ({\rm K-[12]}) \right)
\end{equation}
The Galactic oxygen star sequence (solid curved line) approximately
satisfies an infinite series of the form
\begin{equation}
{\rm J-K} = 13 \times \sum_{n=0}^{\infty} 10^{-3^{n}} {({\rm
K-[12]})}^{3n}
\end{equation}
There is a hint a secondary sequence of oxygen stars, indicated by the
dashed line in Fig.\ 5b. Later we discuss the stars at this side of
the principal sequence in the K--[12] versus J--K diagram in more
detail. From now, we assume that there be no substantial differences
between the K--[12] versus J--K diagrams for mass-losing AGB stars in
the Milky Way and in the LMC.

\subsubsection{LMC sample}

%
%
\begin{figure}[tb]
\centerline{\psfig{figure=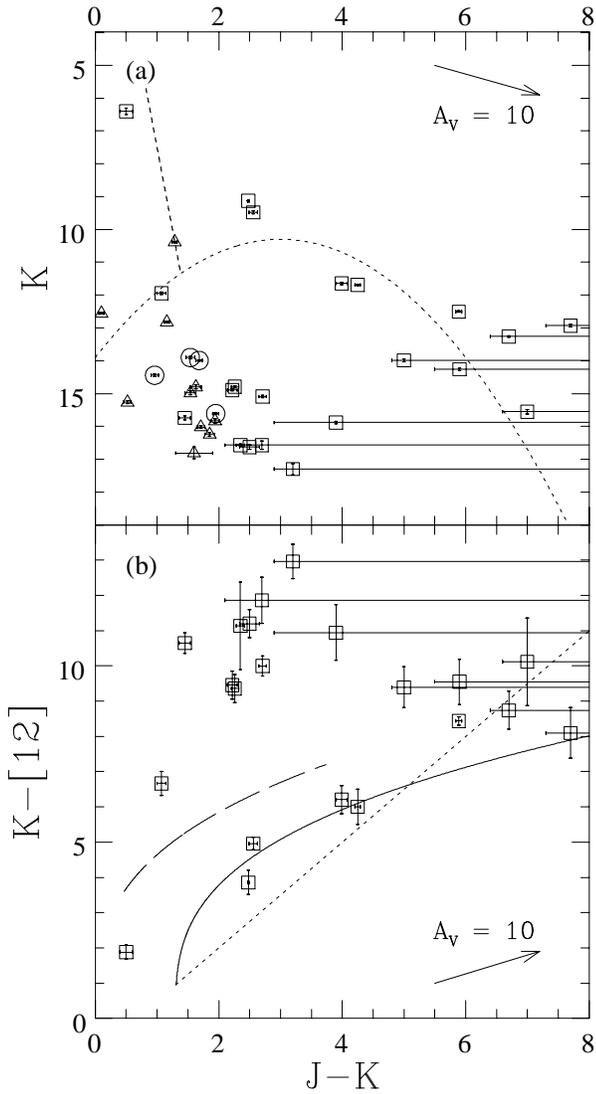,width=90mm}}
\caption[]{K-band magnitudes (a) and K--[12] colours (b) versus J--K
colours for the identified NIR counterparts of the IRAS sources
(squares), the serendipitously detected stars (triangles), and the
detected galaxies (circles), with 1$\sigma$ error bars. We adopt
[12]$ = -2.5 \log{(S_{12}/28.3)}$, with $S_{12}$ the flux density in Jy
in the IRAS 12~$\mu$m band (IRAS Explanatory Supplement 1988). The
dotted lines in the upper panel represent the AGB stars in the LMC from
Loup \& Groenewegen (1994). In the lower panel the Galactic AGB carbon
star (dotted) and oxygen star (solid and dashed) sequences from Fig.\ 5
are indicated. We also plotted arrows corresponding to an extinction in
the visual of A$_V=10$}
\end{figure}

%
%
\begin{figure}[tb]
\centerline{\psfig{figure=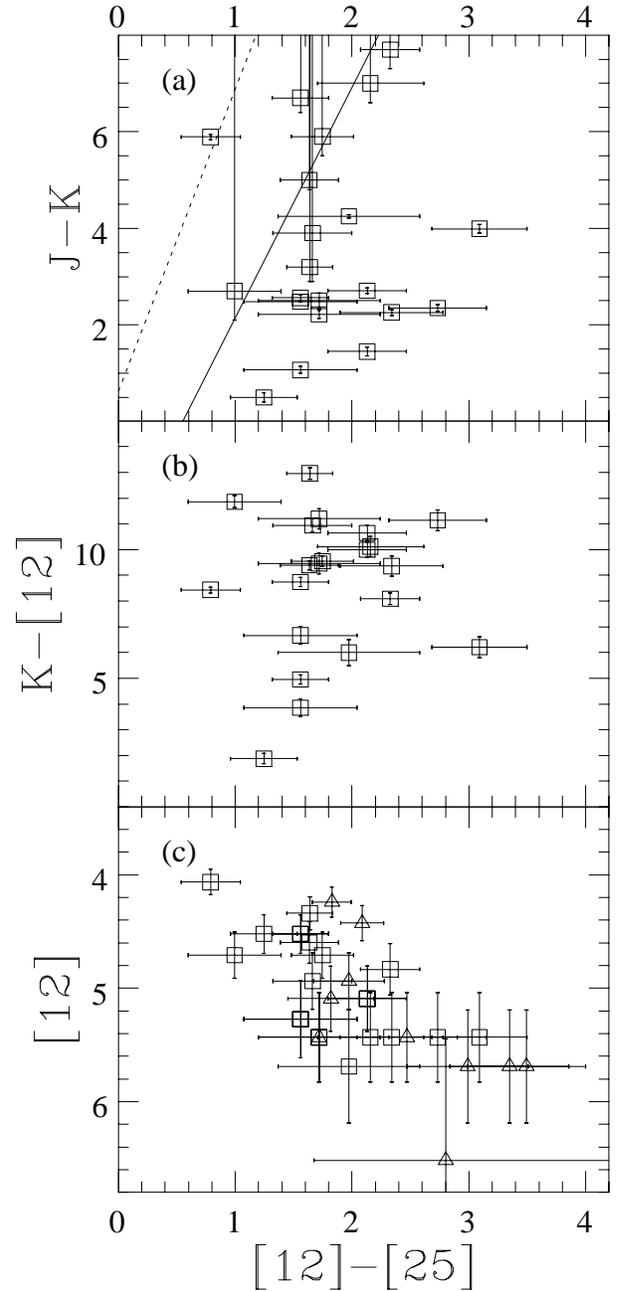,width=90mm}}
\caption[]{J--K (a) and K--[12] (b) colours, and IRAS 12~$\mu$m
magnitudes (c) versus IRAS [12]--[25] colours for the identified NIR
counterparts of the IRAS sources (squares), and the
non-identifications (triangles), with 1-$\sigma$ error bars. We adopt
[12]$ = -2.5 \log{(S_{12}/28.3)}$ and [12]--[25]$ = -2.5
\log{(S_{12}/S_{25})\times(6.73/28.3)}$, with $S_{12}$ and $S_{25}$ the
flux density in Jy in the IRAS 12 and 25~$\mu$m bands, respectively
(IRAS Explanatory Supplement 1988). In the upper panel, the Galactic
AGB carbon star (dotted0 and oxygen star (solid) sequences from Le
Bertre (1993) and Le Sidaner \& Le Bertre (1994) are indicated}
\end{figure}

The K--[12] versus J--K diagram for the stars of the present sample in
the direction of the LMC is given in Fig.\ 6, together with a diagram
of the K-band magnitudes versus J--K. The sequences we have drawn in
the latter diagram are estimated from the positions of AGB stars and
red supergiants in the LMC as compiled by Loup \& Groenewegen
(1994). LMC red supergiants and foreground stars are predominantly
occupying a linear sequence, approximated in Fig.\ 6a by:
\begin{equation}
K = -2.5 + 10 (J-K)
\end{equation}
Mass-losing AGB stars in the LMC follow the curved sequence towards
red J--K colours, approximated in Fig.\ 6a:
\begin{equation}
K = 10.3 + 0.4 \left( (J-K) -3 \right)^2
\end{equation}

Using the K--[12] versus J--K diagram, we can classify several stars
as being mass-losing AGB stars either on the oxygen star sequence
(solid), or on the carbon star sequence (dotted), with the carbon
stars to be found exclusively amongst the optically thickest sources.
But there are several stars that have too large a 12~$\mu$m excess for
their J--K colour, or alternatively are too blue in J--K for their
12~$\mu$m excess to be on the AGB. Of these, LI--LMC1821 is far too
bright in the K-band to be an AGB star in the LMC, and it is probably
a foreground star. Another peculiar source is LI--LMC0530, which lies
on the Loup \& Groenewegen (1994) sequence at a blue J--K colour. We
will come back to this source later. The remaining six outliers are
all faint in the K-band, and five of them are redder than J--K$ = 2$.
Up to three of the stars with lower limits to their J--K colours
might be similar to the group of six outliers as well. The field
stars are distributed over a larger range of K-band magnitudes,
and like the galaxies they are all bluer than J--K$ = 2$. This
suggests that the outliers and the field stars are of different
nature.

Another way of separating AGB carbon stars from AGB oxygen stars is
the J--K versus [12]--[25] colour--colour diagram. For this purpose,
we have estimated the carbon star and oxygen star sequences from Le
Bertre (1993) and Le Sidaner \& Le Bertre (1994). The average carbon
star sequence may be approximated by:
\begin{equation}
[12]-[25] = -0.1 + 0.16 ({\rm J-K})
\end{equation}
The average oxygen star sequence may be approximated by:
\begin{equation}
[12]-[25] = 0.55 + 0.21 ({\rm J-K})
\end{equation}

The J--K versus [12]--[25] diagram for the stars of our sample is
given in Fig.\ 7, together with diagrams of the K--[12] colours and
the 12~$\mu$m magnitudes versus [12]--[25]. For the errors on the IRAS
flux densities we have adopted formal values of 1$\sigma =
0.03$~Jy. The J--K versus [12]--[25] diagram does not work well for
our sample mainly because the lower limits to the J--K colours of
several stars allow these stars to lie either on the oxygen star
sequence (solid), or on the carbon star sequence (dotted). The
outliers of the K--[12] versus J--K diagram have relatively blue J--K
colours. Alternatively they may be characterised by relatively cool
dust envelopes because of their relatively red [12]--[25]
colours. From the [12] versus [12]--[25] diagram it appears that the
IRAS sources for which no counterpart was found (triangles) may have
cooler dust envelopes and/or may experience less severe mass loss, but
these indications are very marginal (see also paper~II). We also note
that if the counterpart would not have been particularly red or bright
as compared to the other stars in the field, we may have failed to
recognise the counterpart as such. This may be the case for PNe, which
can have blue J--K colours between 0 and 0.5 mag due to strong He~{\sc
i} line emission at 1.083~$\mu$m (Whitelock 1985a).

\subsection{Post-AGB star candidates}

It is interesting to compare the identified NIR counterparts of the
IRAS sources as found in the present study, with the sample from van
der Veen et al.\ (1989: VHG-89). The latter sample consists of a
compilation of Galactic objects thought to be in the transition from
the AGB to the planetary nebula phase. Their distances and absolute
magnitudes are not known, so that we can use this sample in the
colour--colour diagrams but not in the colour--magnitude diagrams.

%
%
\begin{figure*}[tb]
\centerline{\psfig{figure=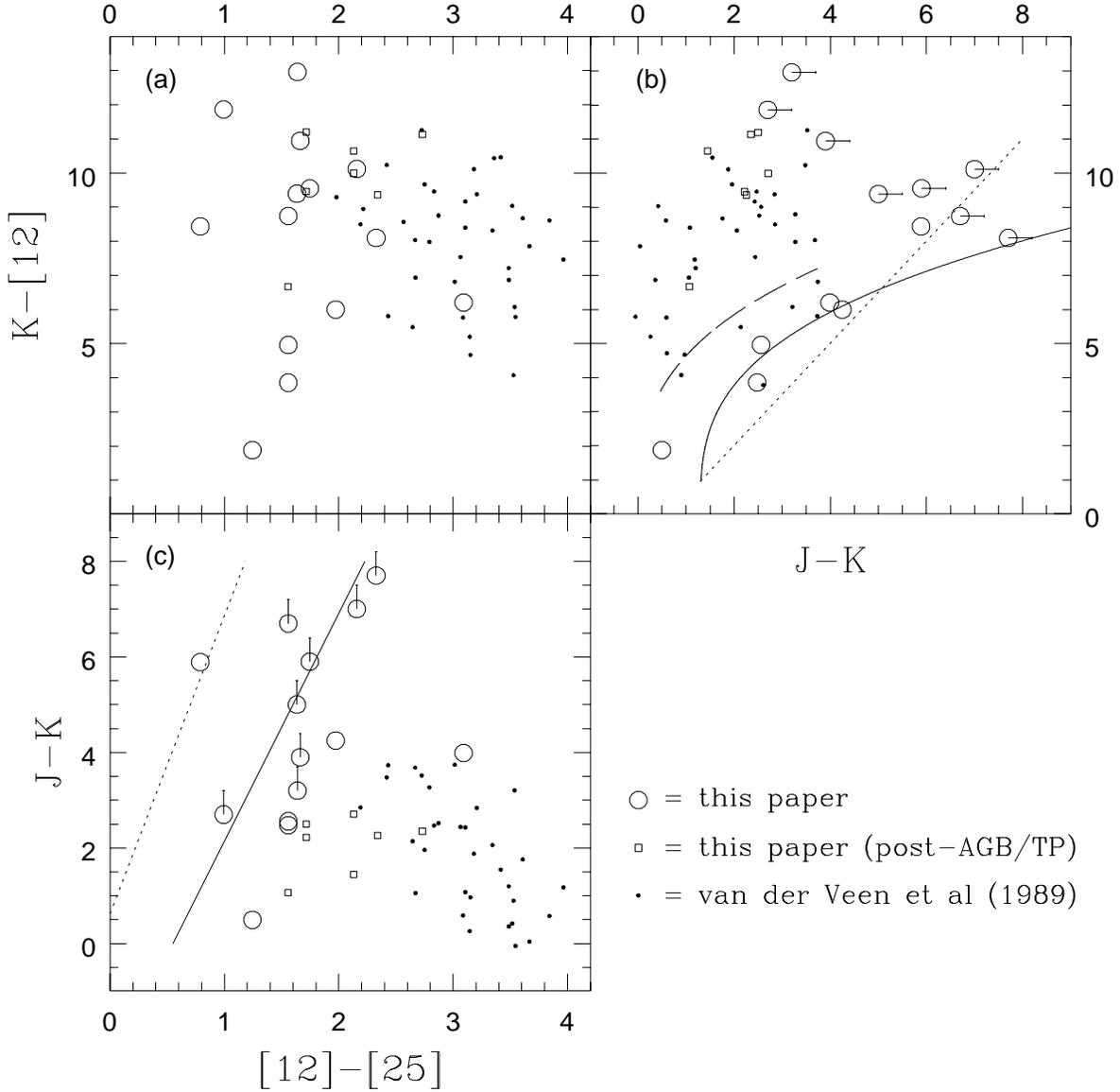,width=160mm}}
\caption[]{K--[12] colours versus [12]--[25] (a) and J--K (b) colours,
and J--K colours versus [12]--[25] colours (c), for the identified NIR
counterparts of the IRAS sources in the present sample (circles, and
squares for the possible post-AGB or Thermal Pulse stars) and the
sample of post-AGB stars of VHG-89 (dots). We adopt [12]$ = -2.5
\log{(S_{12}/28.3)}$ and [12]--[25]$ = -2.5
\log{(S_{12}/S_{25})\times(6.73/28.3)}$, with $S_{12}$ and $S_{25}$ the
flux density in Jy in the IRAS 12 and 25~$\mu$m bands, respectively
(IRAS Explanatory Supplement 1988). The Galactic AGB carbon star
(dotted) and oxygen star (solid and dashed) sequences are as in Fig.\ 6
and Fig.\ 7}
\end{figure*}

The J--K, K--[12], and [12]--[25] colours are plotted versus each
other in Fig.\ 8 (circles for the present sample, and dots for the
VHG-89 sample). The VHG-89 stars have cool dust envelopes, and have
larger 12~$\mu$m excesses for their J--K colours than would be
reconcilable with mass-losing AGB stars. The six outliers of the
present sample roughly overlap with the VHG-89 sample. We classify the
six outliers therefore tentatively as post-AGB stars. They would be
related to the VHG-89 post-AGB stars with relatively warm dust
envelopes, and either have relatively large 12~$\mu$m excesses or be
optically thick in the K-band. We cannot exclude that one or more of
the LMC post-AGB star candidates are actually AGB stars that have
recently experienced a thermal pulse (see section Discussion).

\subsection{The stellar counterpart of LI--LMC1821}

%
%
\begin{table*}
\caption[]{BVRiJK-band magnitudes, and IRAS 12 and 25~$\mu$m flux
densities of the star LI--LMC1821.}
\begin{tabular}{llllllll}
\hline
B & V & R & i & J & K & $S_{12}$ (Jy) & $S_{25}$ (Jy) \\
11.75 $\pm$ 0.15 & 10.27 $\pm$ 0.06 & 9.04 $\pm$ 0.04 & 7.52 $\pm$
0.05 & 6.9 & 6.4 & 0.44 & 0.33 \\
\hline
\end{tabular}
\end{table*}

The IRAS point source LI--LMC1821 is identified with a bright NIR
star. Our BVRi photometry is presented together with the NIR and IRAS
photometry in Table 3. The B--V and V--i colours indicate a
non-reddened, early-M type spectrum (Iyengar \& Parthasarathy
1997). We plot the spectral energy distribution in Fig.\ 9 (squares),
together with an arbitrarily scaled 3000~K blackbody. We also plotted
mean fluxes of the M0 (open circles) and M5 (solid circles) giant from
Fluks et al.\ (1994), scaled to a distance of 1.5 kpc. LI--LMC1821 may
be a late-K or early-M giant at a similar distance, and a member of
the Galactic halo or thick disk population (Robin et al.\
1996). Alternatively, it may be a very luminous red supergiant in the
LMC, with a bolometric luminosity of M$_{\rm bol} \sim -9.7$ mag. The
small IR excess indicates a thin CSE as a result of modest mass
loss. The IRAS colours are consistent with a detached CSE (van der
Veen \& Habing 1988), so that it may be a Galactic post-AGB star, but
as the source is not detected at 60~$\mu$m this is not conclusive. It
is surprising that it has not been listed in any stellar catalogue or
survey, despite that it is bright at optical wavelengths. It would be
interesting to obtain a radial velocity measurement discriminating
between membership of the LMC or the Milky Way.

\section{Bolometric luminosities}

%
%
\begin{figure}[tb]
\centerline{\psfig{figure=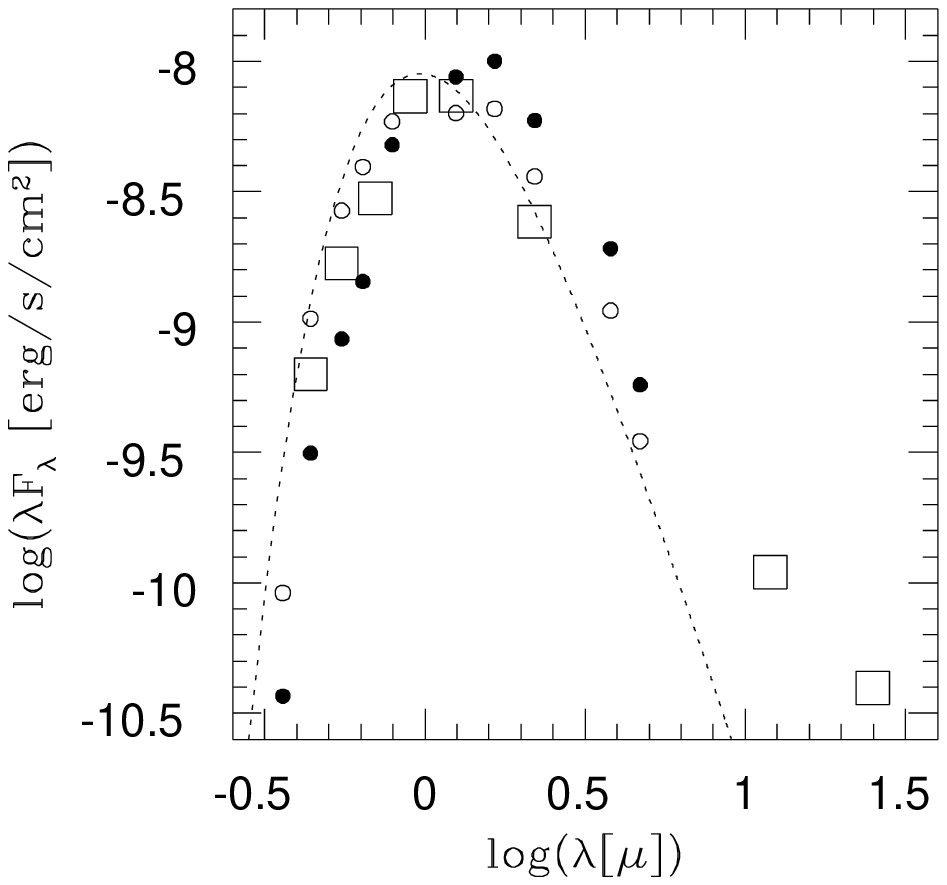,width=90mm}}
\caption[]{Observed spectral energy distribution of LI--LMC1821
(squares), and the M0 (open circles) and M5 (solid circles) giants
from Fluks et al.\ (1994) scaled to a distance of 1.5 kpc, together
with a 3000~K blackbody (dotted)}
\end{figure}

We calculate bolometric luminosities for the stars that we identified
as the NIR counterparts of the IRAS point sources. The method for
estimating the unobserved part of the spectral energy distribution is
based on the average NIR colours for obscured AGB stars, and will be
described in a forthcoming paper. The method for obtaining the
integrated luminosity under the spectral energy distribution is
described in paper~II and originally in Whitelock et al.\ (1994). We
adopted a distance modulus to the LMC of (m--M)$_0 = 18.47$ (Feast \&
Walker 1987). The results are tabulated in Table 4.

The cumulative luminosity distributions of the different classes of
stars are presented in Fig.\ 10a. In principle we could derive the
luminosity distribution by differentiation of the cumulative
distribution function. In practice the errors would be huge because of
the small number of stars. Instead, we apply Gaussian broadening by
replacing each star with absolute bolometric magnitude M$_{{\rm
bol},\star}$ by
\begin{equation}
{\rm n} = {\rm n}_{\rm norm} \times \exp{-({\rm M}_{\rm bol} -
{\rm M}_{{\rm bol},\star})^2}
\end{equation}
where n$_{\rm norm} = 0.564$ to normalise to unity per star. The FWHM
of the star has thereby become 0.83 magnitude. Then we take the sum of
the luminosity-broadened stars. A scientific motivation for applying
the broadening to mass-losing AGB stars is that they are LPVs with
bolometric amplitudes of about one magnitude but only observed at a
single NIR epoch. The luminosity distributions are presented in Fig.\
10b (line types are the same as in Fig.\ 10a). They represent the
distribution over luminosity of the IRAS-detected AGB stars in the
LMC, and they are likely to be incomplete, especially at the faint
end. We assume that this does not seriously affect the relative
distributions of the carbon and oxygen stars. The ratio of the number
of carbon stars over the number of oxygen stars as a function of
luminosity is derived by dividing the luminosity distribution function
of the carbon stars by that of the oxygen stars (Fig.\ 10c). We have
done this for the two extreme cases that all OC stars be carbon stars
(dotted) or that all OC stars be oxygen stars (solid). The real
luminosity relation of the fraction of carbon stars lies in between
these two extremes.

%
%
\begin{table}
\caption[]{Bolometric magnitudes and object classes of the positive
identifications in the LMC: oxygen (O), carbon (C), or inconclusive
(CO) mass-losing AGB stars, and post-AGB or Thermal Pulse (post/TP)
stars.}
\begin{tabular}{llllll}
\hline\hline
LI--LMC & M$_{\rm bol}$ & O & OC & C & post/TP \\
\hline
0099 & --5.42 &          &          & $\times$ &          \\
0109 & --6.28 & $\times$ &          &          &          \\
0136 & --4.95 &          &          & $\times$ &          \\
0180 & --5.01 & $\times$ &          &          &          \\
0297 & --6.36 & $\times$ &          &          &          \\
0326 & --5.27 &          &          &          & $\times$ \\
0344 & --5.82 & $\times$ &          &          &          \\
0530 & --5.55 &          &          &          & $\times$ \\
0603 & --5.65 &          & $\times$ &          &          \\
0777 & --4.99 &          &          &          & $\times$ \\
0782 & --5.50 & $\times$ &          &          &          \\
1092 & --5.57 &          & $\times$ &          &          \\
1198 & --5.30 &          &          & $\times$ &          \\
1316 & --4.83 &          &          &          & $\times$ \\
1624 & --4.82 &          &          &          & $\times$ \\
1721 & --5.27 &          &          &          & $\times$ \\
1803 & --5.13 &          &          &          & $\times$ \\
1813 & --5.78 &          & $\times$ &          &          \\
1817 & --6.12 &          &          & $\times$ &          \\
1818 & --5.89 &          &          & $\times$ &          \\
\hline
\end{tabular}
\end{table}

\section{Discussion}

\subsection{AGB stars}

%
%
\begin{figure}[tb]
\centerline{\psfig{figure=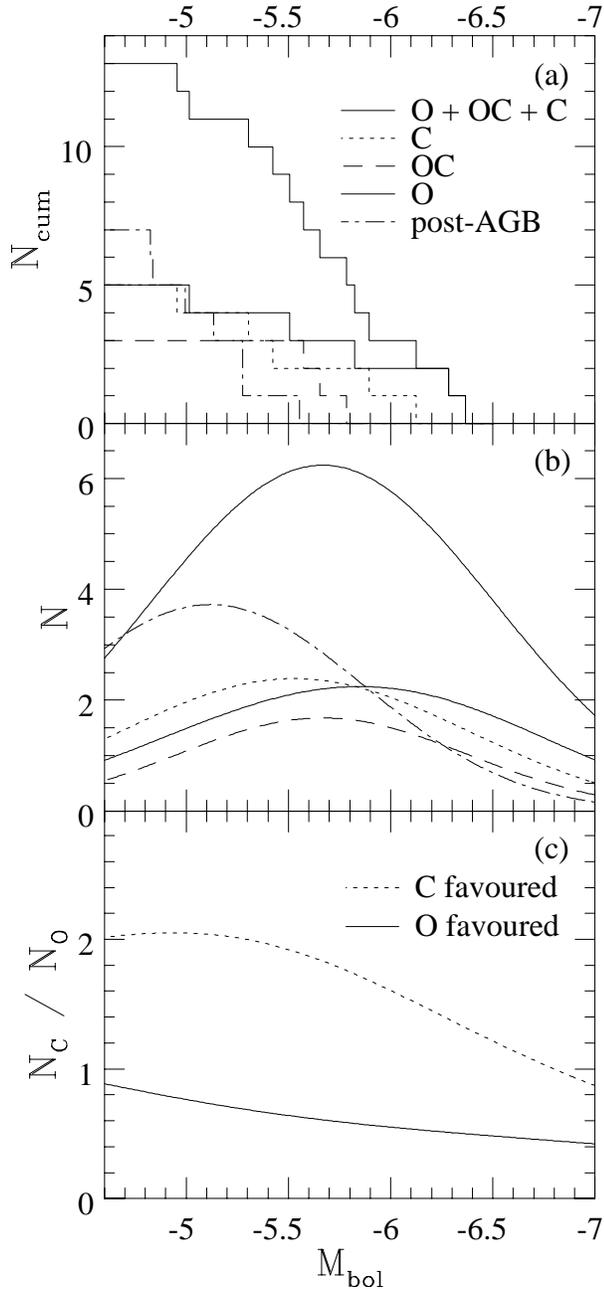,width=90mm}}
\caption[]{(a) Cumulative distribution function of the stars over
absolute bolometric magnitude: ``oxygen'' stars (solid), ``oxygen or
carbon'' stars (dashed), ``carbon'' stars (dotted), all AGB stars
(bold solid), and post-AGB candidates (dash-dotted). (b) Applying
Gaussian broadening (see text), we derive the distribution functions
N. (c) From this we derive the distribution function of the ratio of
carbon to oxygen AGB stars for the two extreme cases of carbon star
favoured (all OC stars are C stars; dotted) and oxygen star favoured
(all OC stars are O stars; solid)}
\end{figure}

The AGB stars that we (re-)discovered have thick CSEs according to
their large J--K and K--[12] colours. Le Sidaner \& Le Bertre (1996)
derive an empirical relation between the continuum$+$line optical
depth $\tau$ of an oxygen-rich CSE at 10~$\mu$m and its J--K colour:
\begin{equation}
\tau \sim 0.8 \times ({\rm J-K}) -1.2
\end{equation}
Apart from the foreground star LI--LMC1821 and the object LI--LMC0530,
the positive identifications between NIR and IRAS sources have J--K
between 2.22 and more than 7.7 mag, indicating optical depths at
10~$\mu$m in the range 0.6---5 or more. All serendipitous detections
have J--K$ < $2 mag, indicating small optical depths. Because we could
not detect all stars in the J-band, the K--[12] colour is a better
tool for studying the optical depth of the obscured AGB stars. Le
Sidaner \& Le Bertre (1996) derive an empirical relation between the
optical depth of the CSE at 10~$\mu$m and the K--[12] colour:
\begin{equation}
\log{\tau} = 4.51 \times \log{({\rm K-[12]})} - 3.37
\end{equation}
Our obscured AGB stars would have optical depths between 0.6 and
45. Post-AGB stars would not satisfy both the J--K and K--[12]
relations for the optical depth, as these two relations define a path
in the K--[12] versus J--K diagram, that is unique to the obscured AGB
stars:
\begin{equation}
({\rm K-[12]})^{4.51} = 1875 \times ({\rm J-K}) - 2813
\end{equation}
It is clear from Fig.\ 11 that Le Sidaner \& Le Bertre give a sequence
for oxygen-rich obscured AGB stars which is reasonably consistent with
our oxygen star sequence. We note that they corrected the IRAS
12~$\mu$m fluxes for the spectral slope, assuming blackbody
temperatures between 300---2000~K. The colour correction factor then
ranges between about 0.92 and 1.38 respectively. This makes their
K--[12] colours change by $-0.09$ mag in the case of a 300~K
blackbody, and 0.35 mag in the case of a 2000~K blackbody. As the
stars with thick CSEs generally have the lower blackbody temperatures,
the difference between the Le Sidaner \& Le Bertre track and our oxygen
star sequence in the K--[12] versus J--K diagram would only be
appreciable at the smallest J--K, and only a few tenths of a magnitude
at most.

The most luminous star in our sample has a bolometric magnitude
M$_{\rm bol} = -6.4$ mag, a factor of two fainter than the theoretical
AGB limit as derived from the Chandrasekhar limit for the core mass
and the core mass-luminosity relation from Paczy\'{n}ski (1971). The
mass-losing AGB star sample of paper~II spanned the luminosity range
between M$_{\rm bol} \sim -4.9$ and $-7.2$ mag. Our new sample
consists of fainter stars on average, but with M$_{\rm bol} = -5.0$
mag our faintest star is not fainter than those of paper~II.

Carbon stars are more numerous at fainter luminosities relative to
oxygen stars. The two brightest stars have oxygen rich CSEs, but the
brightest carbon star, with M$_{\rm bol} = -6.1$ mag, is not much
fainter than these two oxygen stars. On the other hand, the faintest
star is a carbon star, but the faintest oxygen star has the same
M$_{\rm bol} = -5.0$ mag. Hence we conclude that we do not detect a
luminosity regime in which the mass-losing AGB stars are exclusively
either carbon stars or oxygen stars.

%
%
\begin{figure}[tb]
\centerline{\psfig{figure=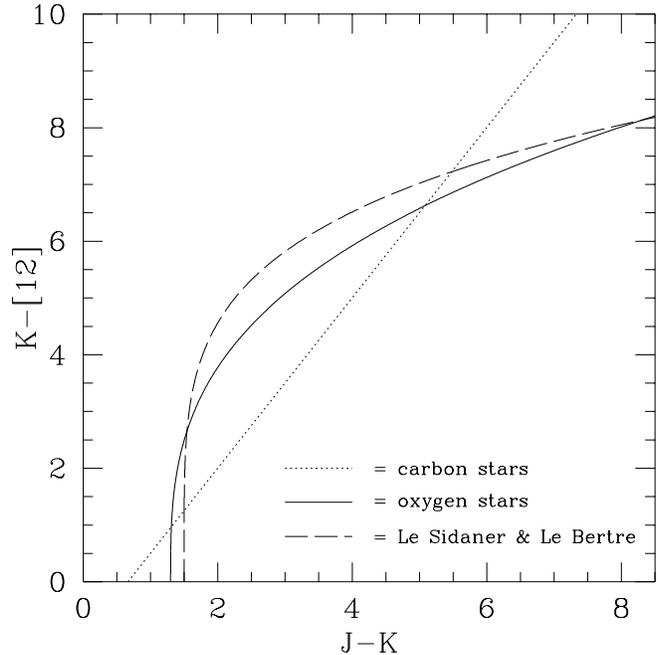,width=90mm}}
\caption[]{K--[12] versus J--K colours according to our sequences for
carbon (dotted) and oxygen (solid) obscured AGB stars, and according
to the parameterisation of the optical depth at 10~$\mu$m by Le
Sidaner \& Le Bertre (1996, dashed). All are derived empirically. We
adopt [12]$ = -2.5 \log{(S_{12}/28.3)}$, with $S_{12}$ the flux density
in Jy in the IRAS 12~$\mu$m band (IRAS Explanatory Supplement 1988)}
\end{figure}

This could mean that luminous AGB stars are prevented from becoming
carbon stars, starting from luminosities as low as M$_{\rm bol} \sim
-5$ mag. Hot Bottom Burning (HBB) may be responsible for this, as the
inner boundary of the convective mantles of the more massive AGB stars
becomes sufficiently hot for CN processing to occur (Sugimoto 1971;
Iben 1975; Scalo et al.\ 1975). The co-existence of oxygen and carbon
stars over a large range of luminosity may be explained as a
consequence of a spread in metallicity. Mass-losing oxygen stars
would be metal poor and experience HBB at M$_{\rm bol} \sim -5$ mag,
whereas mass-losing carbon stars would be more metal rich and not
experience HBB at M$_{\rm bol} \sim -5$ mag. A thorough discussion of
the luminosity distribution function and mass-loss rates of the
mass-losing AGB stars in the LMC is postponed to the next paper in
this series, in combination with the sample of paper~II.

\subsection{Future searches for mass-losing AGB stars}

We have been successful in detecting NIR counterparts in approximately
two out of every three cases. Of these, approximately two out of every
three cases turned out to be mass-losing AGB stars. Of these,
approximately one out of every two cases was too faint to be detected
in the J-band. Would it still be worthwhile to search for NIR
counterparts of the remaining 37 IRAS sources? Current attempts to
find NIR counterparts may in some cases already be limited by the
ground-based NIR searches rather than by the IRAS mid-infrared
detections. The ISO mission is expected to yield an extensive data
base of new mid-infrared point sources. If ISO will detect
bolometrically fainter stars exhibiting similar mass-loss rates to
those of the obscured stars which are IRAS counterparts, then it will
be difficult or impossible to detect their NIR counterparts with
presently available ground-based instruments. However, if it turns out
that (nearly) all of the new ISO detections have NIR counterparts, it
will demonstrate that bolometrically fainter stars exhibit lower
mass-loss rates.

\subsection{Post-AGB stars}

To investigate further the nature of the suspected post-AGB candidates
in our sample, we have taken a closer look at the VHG-89 sample. They
classify their post-AGB candidates into classes {\sc i}, {\sc ii},
{\sc iii}, {\sc iv}a, and {\sc iv}b, based on the shape of the
infrared spectral energy distribution. The higher the class, the more
evolved the post-AGB object: class {\sc i} is characterised by warm
dust completely obscuring the underlying cool star, while in later
classes the dust shell becomes detached, exposing the underlying,
increasingly hotter star. They also argued that classes {\sc i} and
{\sc ii} result from more massive stars than the later classes. For
each class of stars in the VHG-89 sample we have calculated the mean
and standard deviation of the J--K, K--[12], and [12]--[25]
colours. We drew boxes in the colour--colour diagrams, centred at the
mean colours and having sides measuring two times the standard
deviations of the colours. In this way we obtain a schematic picture
of the positions of the VHG-89 classes in the colour--colour diagrams
(Fig.\ 12). Indeed, the dust becomes cooler (larger [12]--[25] colour)
and optically thinner (smaller J--K) with increasing class. The early
classes with obscured stars generally have larger K--[12] and/or
larger J--K than do the optically visible stars from class {\sc iv}.

The LMC post-AGB star candidates could be identified with the VHG-89
classes {\sc ii} and {\sc iii}. The dust of the LMC stars may be
warmer than that of similar stars in the Milky Way if the dust-to-gas
ratio in CSEs of LMC stars is smaller, permitting stellar radiation to
permeate farther out into the CSE, thus heating the dust to higher
temperatures. Also the water abundance may be smaller in the LMC,
causing the oxygen-rich CSEs to cool less efficiently. Similarly,
lower CO and/or HCN abundances may cause warmer carbon-rich CSEs. This
would result in somewhat smaller [12]--[25] in the LMC than in the
Milky Way. The colours of the LMC post-AGB star candidates are indeed
similar to the VHG-89 stars with relatively warm dust. The LMC
post-AGB star candidates may have slightly smaller J--K colours
because their CSEs are optically thinner due to the lower dust-to-gas
ratio, but this could also be explained by selection effects: three of
our newly identified IR stars with large K--[12] colours have lower
limits to their J--K colours that would still permit them to be
post-AGB candidates. On the other hand, the VHG-89 sample is
constructed with a blue cut-off at [12]--[25] = 2 mag, selecting
against the bluemost sources.

%
%
\begin{figure}[tb]
\centerline{\psfig{figure=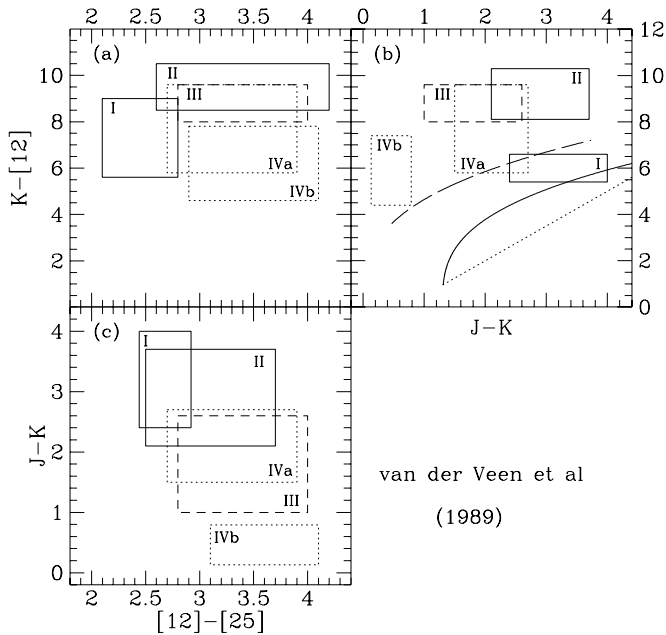,width=90mm}}
\caption[]{K--[12] colours versus [12]--[25] (a) and J--K (b) colours,
and J--K colours versus [12]--[25] colours (c), for the VHG-89 sample
of post-AGB stars. We follow their classification into classes {\sc
i}, {\sc ii}, {\sc iii}, {\sc iv}a, and {\sc iv}b. The boxes are
derived from their data by calculating the mean and standard deviation
for each colour, per object class. We adopt [12]$ = -2.5
\log{(S_{12}/28.3)}$ and [12]--[25]$ = -2.5
\log{(S_{12}/S_{25})\times(6.73/28.3)}$, with $S_{12}$ and $S_{25}$ the
flux density in Jy in the IRAS 12 and 25~$\mu$m bands, respectively
(IRAS Explanatory Supplement 1988). The Galactic AGB carbon star
(dotted) and oxygen star (solid and dashed) sequences are as in Fig.\ 7
and Fig.\ 8}
\end{figure}

We compared the luminosities of the LMC post-AGB star candidates to
those of the five R Coronae Borealis (RCB) stars known in the LMC
(Alcock et al.\ 1996), which are also believed to be post-AGB stars,
and to the luminosities of PNe in the LMC. The latter we took from
Dopita \& Meatheringham (1991), Zijlstra et al.\ (1994), and Dopita et
al.\ (1997). For some of these PNe we know whether they result from an
oxygen- or carbon-rich AGB star. The luminosity distributions are
presented in Fig.\ 13. We note that we do not have complete samples of
any of the type of objects shown. The distributions of the post-AGB
star candidates are only shown down to M$_{\rm bol} = -4.6$ mag, since
this is approximately the detection limit of the combined IRAS-IRAC2b
search. The distributions of the post-AGB stars, RCB stars, and PNe
are very similar, all dropping steeply from M$_{\rm bol} \sim -5$ to
M$_{\rm bol} \sim -6$ mag. In fact, the only object more luminous than
M$_{\rm bol} = -5.6$ mag is the unusual PN SMP-83 (Dopita et al.\
1993).

The PNe for which the C/O ratio is known can be used to derive the
distributions of the oxygen- and carbon-rich PNe in the LMC, and their
ratio (Fig.\ 13c). From this we conclude that the chemical composition
of the PNe is consistent with the chemical composition of the present
small sample of new mass-losing AGB stars.

We note that some of the PNe and two of the RCB stars in the LMC have
been detected by IRAS at 12~$\mu$m at a level of $\sim$0.1---0.2 Jy
(Zijlstra et al.\ 1994; Moshir et al.\ 1992; Alcock et al.\ 1996),
indicating that objects are capable of maintaining their 12~$\mu$m
flux after they have left the AGB. Hence the fact that the LMC
post-AGB star candidates are detected at 12~$\mu$m by IRAS does not
necessarily imply their post-AGB age. Zijlstra et al.\ note that the
IRAS detected LMC PNe have blue [12]--[25] colours, relative to
Galactic PNe. They attribute this to selection effects. The LMC
post-AGB star candidates also have blue [12]--[25] colours, relative
to Galactic post-AGB star candidates. NIR-optical spectroscopy for
the faint, obscured LMC post-AGB star candidates is difficult, but may
be possible. If they are indeed post-AGB stars, their spectra are
expected to be of intermediate type (A, F, G). However, this would not
exclude the possibility that they are binary systems (e.g.\ Whitelock
et al.\ 1995).

If we assume that, down to a certain lower bolometric luminosity
limit we are equally incomplete for the currently available samples of
mass-losing AGB stars (combining paper~II with this paper), post-AGB
stars, and PNe then we can, in principle, estimate their relative
lifetimes. It is more difficult to do this compared to the AGB stars
that have not been detected by IRAS, because many of them will still
evolve significantly in bolometric luminosity. Comparing bolometric
luminosity limited samples of IRAS detected and non-detected AGB stars
therefore results in comparing different populations of stars, with
different main-sequence masses. A synthetic evolution approach is
needed to infer the AGB lifetimes for stars of different main-sequence
masses (Groenewegen \& de Jong 1994). The known PNe have been selected
in a different way than the post-AGB and mass-losing AGB stars,
because of the different observational properties of these
objects. Hence it is not obvious that the currently known sample of
PNe is equally incomplete as the currently known samples of post-AGB
and mass-losing AGB stars. Amongst the IRAS point sources that remain
to be searched for NIR counterparts, there may be a significant number
of post-AGB and/or mass-losing AGB stars, in perhaps different
relative numbers than those presently observed. There may also be
none. Amongst the IRAS point sources that we could not identify with a
NIR counterpart, there may be optically visible post-AGB stars, or
not. The most we can say is that the currently available data suggest
that the mass-losing AGB, post-AGB, and PN stages all have similar
lifetimes, for a star with an AGB-tip bolometric luminosities between
M$_{\rm bol} \sim -5$ and $-6$ mag, i.e.\ that has a progenitor mass
in the range 2.5--4 M$_{\odot}$ (Vassiliadis \& Wood 1993). More
massive stars seem to have post-AGB and PN lifetimes that are
considerably shorter than their mass-losing AGB lifetimes.

\subsection{Thermal Pulse stars}

%
%
\begin{figure}[tb]
\centerline{\psfig{figure=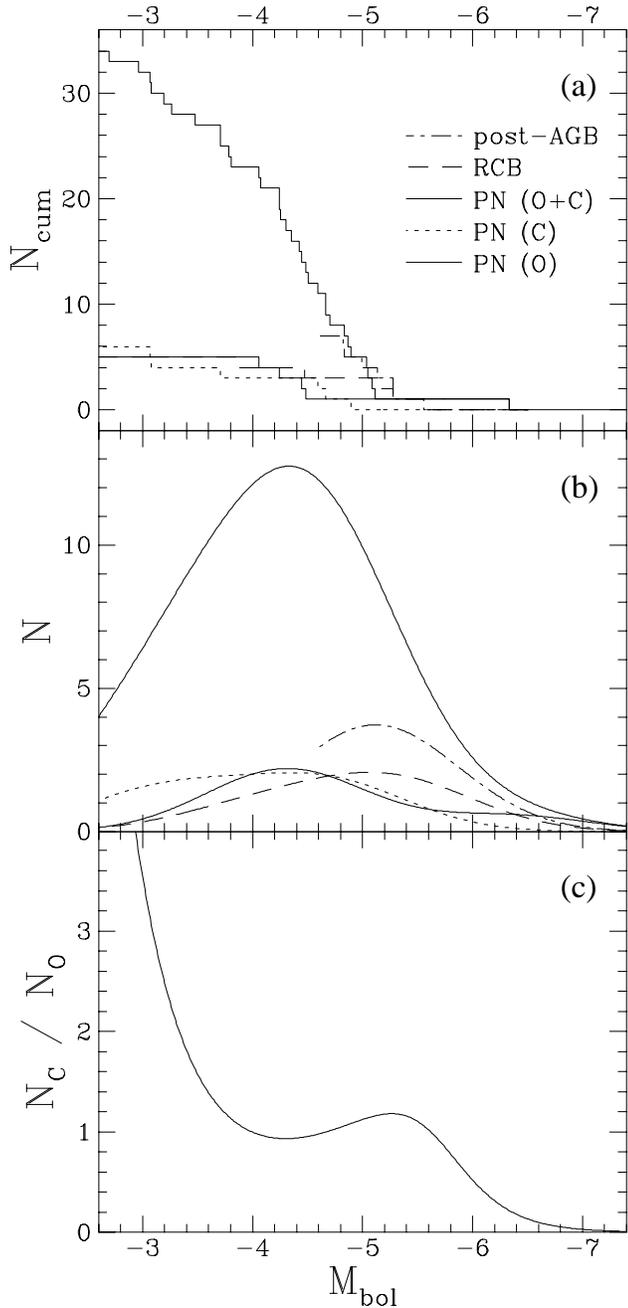,width=90mm}}
\caption[]{(a) Cumulative distribution function of the LMC objects
over absolute bolometric magnitude: post-AGB star candidates
(dot-dashed), R Corona Borealis stars (dashed), oxygen-rich PNe
(solid), carbon-rich PNe (dotted), and all PNe (including the PNe for
which the C/O ratio is not known; bold solid). (b) The distribution
functions derived in the same way as in Fig.\ 10. (c) The distribution
function of the ratio of carbon-rich to oxygen-rich PNe}
\end{figure}

The suspected secondary sequence in the K--[12] versus J--K diagram
for Galactic oxygen stars does have some overlap with the VHG-89
stars, but the latter are mostly found having larger K--[12]. This can
be partly due to larger optical depths of the CSEs of the VHG-89
stars, to the extent that the CSE becomes optically thick in the
K-band, but it cannot explain the position of the VHG-89 class {\sc
iv} at small J--K. Probably the VHG-89 stars have more massive CSEs,
yielding a larger 12~$\mu$m flux, but these CSEs are detached,
yielding small column densities towards the star and hence small
J--K. Consequently, stars at the secondary sequence are not expected
to evolve directly into stars with larger K--[12]. If the stars from
VHG-89 are post-AGB stars, then they are expected to have evolved from
large J--K and K--[12], first getting smaller J--K before also getting
smaller K--[12]. Hence the secondary sequence may be related to the
VHG-89 class {\sc i} only. The fact that we noticed a secondary
sequence, rather than a gradient from the primary sequence into the
VHG-89 areas of the K--[12] versus J--K diagram, suggests that the
secondary sequence and perhaps the VHG-89 class {\sc i} are not
related to the VHG-89 post-AGB classes {\sc ii}---{\sc iv}. We can
explain them instead as stars that only temporarily stopped losing
mass, possibly as a result of a thermal pulse (TP). They are then
expected to return to the AGB and resume heavy mass loss (Zijlstra et
al.\ 1992).

It is difficult to discern a TP star from a post-AGB star, since both
experience the same phenomenon: the CSE becomes detached. But at least
statistically there are differences to be expected, that can be
observed in the near- and mid-IR: the K--[12] is expected to be
statistically larger for a post-AGB star than for a TP star. There are
three reasons for this. First, post-AGB stars shrink while maintaining
the same bolometric luminosity, and therefore they must increase in
effective temperature. This results in increased heating of the CSE,
counteracting at least partly the cooling of the CSE as a result of
its increasing distance to the star. TP stars do not change as much in
effective temperature, and hence the detaching CSE of a TP star cools
more rapidly than that of a post-AGB star. Second, as the effective
temperature increases but not the bolometric luminosity, the star will
become fainter in the K-band, increasing its K--[12] colour. Although
TP stars may have decreased in bolometric luminosity, this is expected
to be at most one magnitude (Vassiliadis \& Wood 1993). Third,
post-AGB stars are more evolved than TP stars. They have experienced
mass loss for a longer time span, yielding more massive CSEs and
consequently larger 12~$\mu$m fluxes.

LI--LMC0530 has too small a [12]--[25] colour for its K--[12] colour
to be a normal post-AGB star candidate, relative to both the VHG-89
stars and the six LMC post-AGB star candidates discussed above. It is
also much brighter in the K-band than the other LMC post-AGB star
candidates, which suggests an optically thin CSE. Whitelock et al.\
(1995) found two stars in the South Galactic Cap that they explained
as TP stars, because of their lack of variability. These two stars
have J--K and [12]--[25] colours similar to those of LI--LMC0530, but
their K--[12] colours are much smaller than that of LI--LMC0530, and
even smaller than the secondary sequence in the K--[12] versus J--K
diagram. In fact these Galactic TP star candidates are more similar to
normal mass-losing AGB stars. This can be understood if they
experienced one of their first TPs, before they had built up a massive
CSE that yields a significant 12~$\mu$m excess. LI--LMC0530 seems to
take an intermediate position between post-AGB stars and TP stars.

LI--LMC0530 is identified with the LPV SHV0510004--692755: an LPV with
an I-band amplitude of 1.24 mag and a period of 169 days. Hughes \&
Wood (1990) give a spectral type of M6. It obeys the period-luminosity
relation for the K-band magnitude perfectly, although bolometrically
it is 4---5 times more luminous than the AGB stars that define the
period-luminosity relation (Reid et al.\ 1995). LI--LMC0530 must have
been a mass-losing AGB star in the past, because it has a large
12~$\mu$m excess. It may now be pulsating in a (high) overtone, and
maybe it will resume Mira variability and mass loss. If it is a
post-AGB star at present, then it must be in a very early post-AGB
stage, because its effective temperature has not increased much. But
in this case we would expect the star to be obscured by an optically
thick CSE. This contradicts its small J--K, unless we invoke a highly
aspherical shape of the CSE. We conclude that the nature of
LI--LMC0530 is uncertain: it may be an early post-AGB star of class
{\sc iv}, or a star recovering from the effects of one of its last
thermal pulses on the AGB. Alternatively, the measured flux at
wavelengths shorter than $\sim 2 \mu$m may have been affected by the
presence of another star in the line-of-sight.

\section{Summary}

Our search for NIR counterparts of IRAS point sources in the direction
of the LMC was aimed at finding obscured AGB stars in the LMC: we
found 13 of them. The K--[12] versus J--K diagram is shown to be well
suited for discriminating between mass-losing oxygen and carbon
stars. Comparison between the luminosity distributions of the oxygen
and carbon stars suggests that the fraction of carbon stars is smaller
at bolometric luminosities around M$_{\rm bol} \sim -6$ mag than
around M$_{\rm bol} \sim -5$ mag. Oxygen and carbon stars co-exist at
all luminosities, but the new sample of obscured AGB stars does not
include luminosities as bright as M$_{\rm bol} \sim -7$ mag. Analysis
of the combined sample of known obscured AGB stars in the LMC will be
presented in the next paper of this series.

Besides the obscured AGB stars, we found a probable Galactic halo star
or very luminous LMC red supergiant. We also found 7 probable IRAS
counterparts which have large J--K colours, but not as large as the
obscured AGB stars. Comparison with a sample of post-AGB objects in
the Milky Way as compiled by van der Veen et al.\ (1989) suggests that
these NIR detections are likely to be post-AGB stars, with
luminosities around M$_{\rm bol} \sim -5$ mag. We investigated the
possibility that they are AGB stars recovering from the occurrence of a
thermal pulse. Although this cannot be ruled out completely, we found
it difficult to reconcile their NIR and IRAS colours with those of
stars in the Milky Way that are suspected to experience this short
evolutionary phase.

We also detected several objects with J--K colours larger than for
normal stars, but not as large as the IRAS counterparts. They are not
related to the IRAS source in the field. Although two of them are
identified with known LPVs in the LMC, the remaining few stars may be
LMC stars that are reddened by interstellar extinction inside the
LMC. This suggestion is supported by the serendipitous detection of a
few galaxies with larger J--K colours than normal galaxies, indicating
reddening by the LMC corresponding to a visual extinction of a few
mag. This would have severe consequences for the study of stars inside
the LMC, that are often assumed not to suffer significant interstellar
extinction.

\acknowledgements{We would like to thank Drs.\ Montserrat
Villar-Mart\'{i}n, Rene Mendez, Christopher Lidman, Martin
Groenewegen, and Pierre-Alain Duc for discussion on various aspects of
our research. We thank the referee Dr.\ Jeremy Mould for his comments
which helped improving the paper. We acknowledge the allocation of
Director's Discretionary Time at the Dutch telescope at ESO/La Silla.
We made use of the SIMBAD database, operated at CDS, Strasbourg,
France. Jacco queria expresar sus mejores gracias a Montse por
haberle contestado {\em si} a una pregunta bien profunda.}

\end{document}